\documentclass[useAMS,usenatbib]{mn2e}
\usepackage{graphicx}
\usepackage{graphicx,color}
\usepackage{textcomp}
\usepackage{amsmath}
\usepackage{hyperref}
\usepackage{aecompl}
\usepackage[T1]{fontenc} 
\DeclareGraphicsExtensions{.eps, .ps}

\newcommand{\mnras}{MNRAS}
\newcommand{\aap}{A\&A}
\title[Stellar kinematics and dynamical model for NGC\,4258.]{The benchmark
black hole in NGC\,4258: dynamical models from high-resolution two-dimensional
stellar kinematics}

\author[Drehmer et al]{Daniel
Alf Drehmer$^{1,2}$\thanks{E-mail:alfdrehmer@gmail.com}, Thaisa Storchi
Bergmann$^{1}$,  Fabricio Ferrari$^{3}$,\and Michele Cappellari$^{4}$ and
Rogemar A. Riffel$^{5}$\\
$^{1}$Universidade Federal do Rio Grande do Sul, Instituto de F\'isica, CP 15051, Porto Alegre 91501-970, RS, Brasil.\\
$^{2}$Universidade Estadual de Maring\'a, Departamento de Tecnologia, 87506-370 Umuarama, PR, Brasil.\\
$^{3}$Universidade Federal de Rio Grande, Instituto de Matem\'atica
Estat\'istica e F\'isica, CP 474, 96201-900, Rio Grande, RS, Brasil.\\
$^{4}$Sub-department of Astrophysics, Department of Physics, University of
Oxford, Denys Wilkinson Building, Keble Road, Oxford OX1 3RH, UK.\\
$^5$Universidade Federal de Santa Maria, Departamento de F\'i­sica, Centro de
Ci\^encias Naturais e Ex\'atas, 97105-900, Santa Maria, RS, Brasil.}

\begin{document}

\date{Accepted, 9 March 2015. Received, 5 March 2015; in original form, 5 August
2014}

\pagerange{\pageref{firstpage}--\pageref{lastpage}} \pubyear{2015}

\maketitle

\label{firstpage}

\begin{abstract}
NGC\,4258 is the galaxy with the most accurate (maser-based) determination for
the mass of the supermassive black hole (SMBH) in its nucleus.
In this work we present a two-dimensional mapping of the stellar kinematics in
the inner 3.0\,$\times$\,3.0 arcsec = 100\,$\times$\,100\,pc
of NGC\,4258 using adaptative-optics observations obtained with the Near-Infrared
Integral Field Spectrograph  of the GEMINI North telescope at a $\approx$\,0.11 
arcsec (4\,pc) angular resolution. The observations resolve the radius of influence 
of the SMBH, revealing an abrupt increase in the stellar velocity dispersion within 
$\approx$\,10\,pc from the nucleus, consistent with the presence of a SMBH there. 
Assuming that the galaxy nucleus is in a steady
state and that the  velocity dispersion ellipsoid is aligned  with  a
cylindrical coordinate system, we constructed a Jeans anisotropic dynamical 
model  to  fit  the  observed kinematics distribution. Our  dynamical  model 
assumes  that  the galaxy has axial symmetry and is constructed using the
multi-gaussian expansion method to parametrize  the observed  surface 
brightness distribution. The Jeans dynamical model has three free parameters:
the mass of the central SMBH ($M_\bullet$), the mass-luminosity ratio
($\Gamma_k=M/L$) of the galaxy  and the anisotropy of the velocity
distribution. We test two types of models: one with constant velocity
anisotropy, and another with variable anisotropy. The model that best
reproduces the observed kinematics was obtained considering that the galaxy
has radially varying anisotropy, being the best-fitting parameters  with
3$\sigma$ significance $M_\bullet=4.8^{+0.8}_{-0.9}\times 10^7\,{\rm M_\odot}$
and $\Gamma_k = 4.1^{+0.4}_{-0.5}$. This value for the mass of the
SMBH is just 25 per cent larger than that of the maser determination
and 50 per cent larger that a previous stellar dynamical determination obtained
via Schwarzschild models for long-slit data that provides a SMBH mass 15 per
cent lower than the maser value.

\end{abstract}


\section{Introduction}\label{sec:intro}

NGC\,4258 (M\,106) is a spiral galaxy with Hubble
type SABbc, at a distance of 7.28$\pm$0.3\,Mpc \citep{herrnstein99},
which harbours one of the closest active galactic nucleus (AGN), classified as
Seyfert 1.9. This galaxy is well known from previous studies for
harbouring the supermassive black hole (SMBH) with the best constrained
mass after that of the Milky Way, with a value of
$M_{\bullet\,\rm Maser}=(3.82\pm0.2)\times10^{7}\,{\rm M_\odot}$, obtained from
resolved kinematics of a rotating H$_2$0 maser disc within 0.13\,pc from
the nucleus \citep{miyoshi95, greenhill95, herrnstein99}. It is also
well known for its anomalous arms, which resemble spiral arms but are more
diffuse than regular spiral arms  \citep{wilson01}. The anomalous arms span
5\,arcmin in optical line emission and 10\,arcmin in radio continuum emission
\citep{cecil00}.

A nuclear non-stellar continuum and broad optical emission lines have been
observed in  polarized light \citep{wilkes95}, supporting the presence of an
obscured AGN, from which a radio jet is observed propagating
perpendicularly to the maser disc \citep{cecil92}. The jet is oriented
at a position angle (hereafter PA) PA=$-3^\circ\pm1^\circ$, consistent
with the projected spin axis orientation of the maser disc, which has a
major axis PA=86$^\circ\pm2^\circ$ \citep{cecil00}. Farther from the
nucleus, interactions between the radio jet and the interstellar gas are
probably the origin of the anomalous arms \citep[][and references
therein]{wilson01}.

In the infrared (hereafter IR), the nuclear continuum has been found to
be well reproduced by a power-law $f_\nu\propto\nu^{-1.4}$, which seems
to extend through the optical to the ultraviolet \citep{yuan02}. In
X-rays, the nuclear spectrum presents several  components: two
power-laws, a thermal component and the
Fe\,K$\alpha$ emission-line \citep{yang07}. The nuclear source is not
resolved in the IR (1 to $18{\rm \mu m}$ spectral region) in observations with
$0.2\,{\rm arcsec}$ angular resolution and presents total luminosity of
$2\times10^{8}\,{\rm L_\odot}$ \citep{chary00}.

The host galaxy disc has the photometric major axis at PA=$150^\circ$, as
obtained from H\,{\sc i} observations \citep{vanAlbada80}, who have also
obtained an inclination for the disc of $i=72^\circ$. A previous H$\alpha$ study
by \citet{vanderkruit74} gave values of PA=146$^\circ$ and $i=64^\circ$. More
recently \citet{sawada-satoh07} have mapped the large scale molecular
CO(2-1) velocity field and concluded it is dominated by rotation in a
disc with major axis PA=$160^\circ$ and $i=65.6^\circ$.

Other studies include the one of \citet{pastorini07}, who measured the gas
kinematics close to the nucleus using long-slit spectra obtained with the Space
Telescope Imaging Spectrograph ({\it HST}-STIS) and that of
\citet{herrnstein05}, who constrained even further the SMBH mass using new maser
observations. There is also a more recent study by \citet{siopis09} which uses
the long-slit spectra from {\it HST}-STIS to obtain the stellar velocity
distribution
using the Ca\,{\sc ii} triplet absorption lines from $2$ to $18.2 \,{\rm
arcsec}$ along the major axis and to $11.7\,{\rm arcsec}$ along the minor axis
of the galaxy, obtaining a direct determination of the mass of the SMBH.
  
Adopting the SMBH mass obtained from the kinematics of the maser disc, and
using a velocity dispersion for the galaxy bulge of $105\,{\rm
km\,s^{-1}}$\footnote{This is the average value of the velocity
dispersion in the whole NIFS field of view and not the luminosity weighted
velocity dispersion inside a single aperture that contains the whole galaxy
bulge (see Sec.\,\ref{sec:losvd}).}, one can calculate the 
radius of influence of the SMBH $\approx$15\,pc. For a scale of  35\,pc per 
arcsecond, this  radius corresponds to 0.42\,arcsec, thus resolvable with 
ground-based adaptative optics observations, that typically reach 0.1\,arcsec in the 
near-infrared. We thus decided to observe NGC\,4258 with the Gemini 
Near-Infrared Integral Field Spectrograph (NIFS) in order to map the stellar 
kinematics and investigate the effect of the presence of the SMBH in the inner 
dynamics of the galaxy. Although we also use stellar kinematics to probe the 
presence of the SMBH as done by \citet{siopis09}, our \textit{K}-band data are 
less affected by reddening, allowing the mapping of stellar kinematics
much closer to the nucleus than previous observations. In addition, our NIFS
data provide a two-dimensional coverage instead of the limited spatial coverage of
long-slit observations.

Although the use of dynamical models based on the Schwarzschild orbit
superposition method \citep{Schwarzschild1979} is currently the most popular
method to determine the masses of SMBHs in the centre of galaxies, the
Jeans dynamical models have proved capable of reproducing the values of the
masses of the SMBH in good agreement with those determined by the
Schwarzschild models \citep{Cappellari2010, krajnovic2009}. The Jeans models
are also  useful when one wants to have more physical insight to the problem. In
addition, a Schwarzschild orbit superposition model of NGC\,4258 already exists
\citep{siopis09}, providing a value for the mass of the SMBH of
$M_{\bullet\,{\rm Schw}} = (3.3\, \pm 0.2) \times 10^7{\rm M_\odot}$,
which is comparable to the  better constrained value ($M_{\bullet \rm
Maser} = (3.82 \pm 0.02) \times 10^7{\rm M_\odot}$)  derived from maser
observations
\cite{herrnstein05}.

We have thus decided to study the dynamics of the nuclear region of NGC\,4258
using our NIFS data  and Jeans models in order to determine the dynamical
quantities that govern the system. Our goal is also to investigate how well
these models reproduce the observed stellar kinematics and how large are the
differences between the values of the mass of the SMBH of NGC\,4258 determined
by different methods. The use of the Jeans anisotropic
dynamical model has allowed us to understand how variations in the model
parameters -- such as the velocity anisotropy and the mass-to-light ratio -- 
affect the stellar kinematics and the determination of the mass of the SMBH in
NGC\,4258.
\begin{figure}
  \centering
    \includegraphics[width=.5\textwidth]{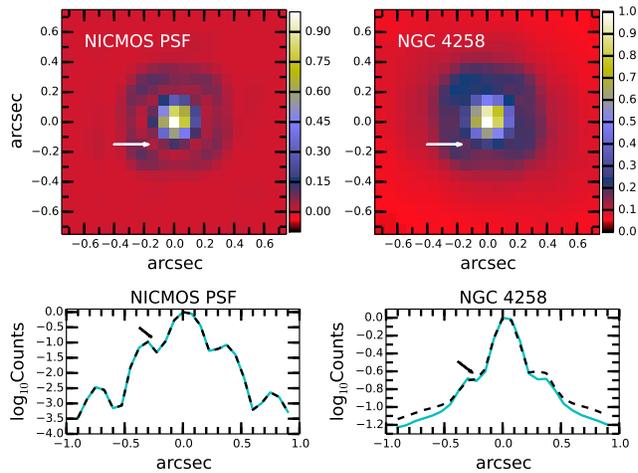}
  \caption{Top Left: the NICMOS PSF modelled with the {\sc tiny tim}
software. Top Right: the central $1.5\,{\rm arcsec}$ of the NICMOS image of the
galaxy presenting the the features of the NICMOS PSF. Intensity units are
normalized counts. The bottom panels show cuts along the two diagonals (dashed
line and cyan continuum line) across the upper panels, in
logarithmic scale.}\label{fig:psf}
\end{figure}
\begin{figure}
  \centering
  \includegraphics[width=.45\textwidth]{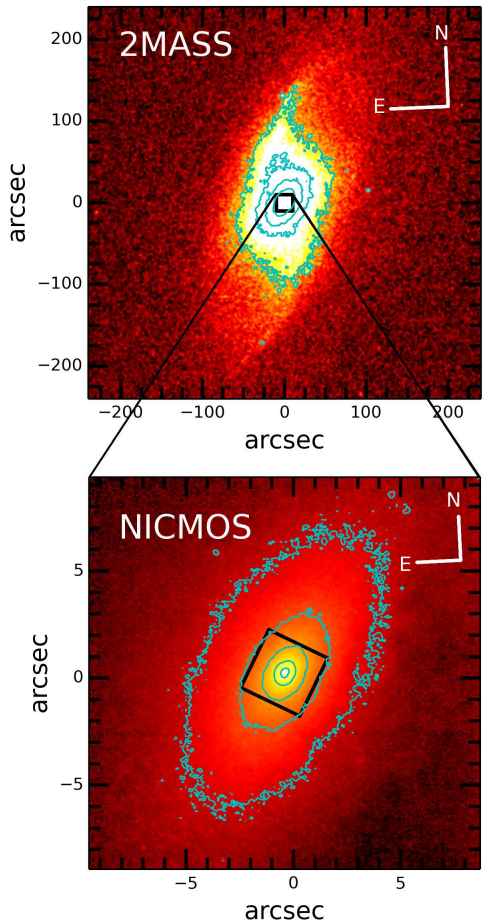}
\caption{Top: the 2MASS \textit{Ks}-band image. Bottom: the \textit{K}-band
image from NICMOS. The rectangle in the bottom panel shows the
field of view of our NIFS observations presented in Fig
\ref{fig:espectro}.}\label{fig:photo}
\end{figure}

This paper is organized as follows. In section \ref{sec:data} we
present our IFU observations and the photometric data used to construct the
dynamical model. In Section \ref{sec:losvd} we discuss the derivation of the
two-dimensional line of sight velocity distribution (LOSVD). Section
\ref{sec:model} is dedicated to the description of the dynamical model and
the discussion of the results and in section \ref{sec:conclusion}  we present
our concluding remarks.

\section{Observations and data reduction}\label{sec:data}

The Integral Field spectroscopic data were obtained in the near-IR
\textit{K}-band with the instrument NIFS \citep{mcgregor03} operating with the
ALTAIR adaptive optics module at the 8-m Gemini North telescope in April 2007
under the science program GN-2007A-Q-25. The \textit{K}-band was selected for
this study because it contains the absorption CO bands at $\approx 2.3\,{\rm \mu
m}$ which can be used to derive the stellar kinematics. In order to obtain the
surface brightness distribution to model the galaxy mass density distribution
(and the gravitational potential) we used two sets of archive images: the first
of them obtained with the Near-Infrared Camera and Multi Object Spectrometer
(NICMOS) NIC2 camera of the {\it HST}. The second is a large scale image from
the Two Micron All Sky Survey (2MASS). In the next sections we describe these
sets of data and the reduction procedures.

\subsection{The photometric data}\label{sec:photodata}

\begin{figure*}
\centering
\includegraphics[width=1.00\textwidth]{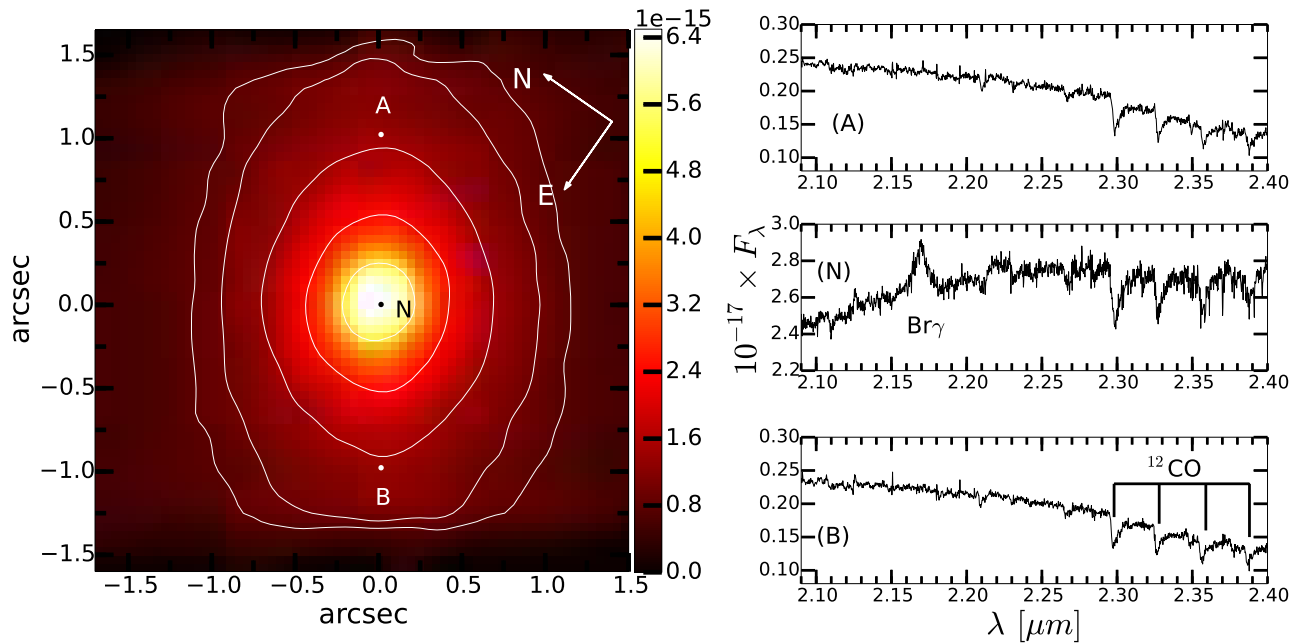}
\caption{Left: Continuum map (reconstructed image) from our NIFS spectroscopy.
Right: Spectra at the positions A, N and B marked at the left panel with
the Br$\gamma$ emission-line and CO absorption bands identified. Flux units
are erg\,s$^{-1}$\,cm$^{-2}$ and the extraction apertures are 0.05 arcsec x 0.05
arcsec.} 
\label{fig:espectro}
\end{figure*}

\subsubsection{NICMOS}

Near-infrared \textit{K}-band images were obtained from the Multimission Archive
at STScI (MAST) corresponding to data sets N46801050 for object observations and
N46801060 for sky observations, from the proposal ID 7230
\citep{scoville97}. These observations were obtained with the NIC2 camera of
NICMOS using the F222M filter. The NIC2 camera has $256\times256$ pixels, with
scales of $0.075\,{\rm arcsec\,pix^{-1}}$ in $x$ and $y$, providing a field of
view of $\sim\,19.2\times19.2\,{\rm arcsec^2}$. The F222M filter covers
the wavelength range $2.15-2.28\,{\rm \mu m}$, with central wavelength at
$2.2174\,{\rm \mu m}$. The galaxy observation have a total exposure time of
$224$ seconds performed in {\sc multiaccum} mode with a four point spiral
dither; the sky observations were obtained with a three point spiral dither.

The data reduction was done with standard {\sc iraf}\footnote{IRAF is distributed
by the National Optical Astronomy Observatories, which are operated by the 
Association of Universities for Research in Astronomy, Inc., under cooperative 
agreement with the National Science Foundation.}  and {\sc pyraf} tasks following
the same steps as in Chary et. al. (2000). We used the
{\sc multidrizzle} task to combine the dithered images following the standard
sequence for NICMOS dithered observations described in the {\sc multidrizzle}
handbook. Inspecting the data quality files we noted that due the 
emission of the AGN, some of the pixels at the centre of the galaxy are incorrectly
identified as cosmic rays in the four  dithered images. We thus corrected this
in the data quality files and created a static mask to correct for a spurious
``bar'' feature  in the column 128. Then a separated drizzled image was created
for each dither position of the detector. To determine the need for extra
offsets we have done cross correlation measurements between the images with the
{\sc crossdriz} task and the extra shifts have them been computed with
the  {\sc shiftfind} task. Then we ran again the {\sc multidrizzle} with the
extra shifts and created new separately drizzled images as well as a
well-aligned median image. Finally, we ran {\sc multidrizzle} again to derive
the cosmic ray masks and combine the images in a final drizzled image, clean of
cosmic rays and detector artifacts.

The compact nuclear emission due to the AGN its evident in the reduced image.
This unresolved emission appears in the image as a point source introducing
the spurious features of the NICMOS point spread function (PSF) in the image. In
the top left panel of Fig.\,\ref{fig:psf} we show the NICMOS PSF, and in the
right panel the central $1.5\,{\rm arcsec}$ of the image of NGC\,4258.
In the bottom panels of this figure we display cuts along the two
diagonals of each of the the images in the top panels (black dashed and cyan
continuous lines). The arrows indicate the strong signature of the PSF in the
image of the galaxy. As our interest is to determine the luminosity distribution
and gravitational potential generated by the stars, we need to subtract the
contribution from the AGN. In order to do this, we generated a large image of
the NICMOS PSF and centred it at the same position of the galaxy centre in the
detector. We then normalized the PSF image so that its peak flux coincided with
that of the galaxy. After several tests in which we multiplied the normalized
PSF by different fractions and subtracted it from the galaxy image, we concluded
that the smoothest residual (without an ``inverted peak'' due to
over-subtraction) was obtained with a contribution of the PSF of 70 per cent to
the
peak of the brightness distribution. The resulting image after reduction and
AGN subtraction is shown in the bottom panel of Figure \ref{fig:photo}. The
rectangle in the centre of the image represents the field of view of the NIFS
observations. The total amounth of light subracted corresponds to
$\approx$\,1.9\,$\times$10$^7\,{\rm L_{\odot K}}$. In Apendix\,
\ref{app:oversubtraction} we perform some tests to evaluate the effects of a
possible over- or under-subtraction of the AGN on the determination of the mass
of the SMBH.

\subsubsection{2MASS}
In order to characterize the surface brightness distribution of the galaxy
beyond the inner $19.0\times19.0\,{\rm arcsec^2}$ we used a large scale image
from the 2MASS telescope in the \textit{K}-short-band (\textit{Ks}-band) from
\citet{jarrett03}. The 2MASS image has a scale of $1.0\,{\rm arcsec\,pix^{-1}}$
and the \textit{Ks}-filter covers the wavelength
range from 1.93 to $2.38\,{\rm \mu m}$ with central wavelength of $2.17\,{\rm
\mu m}$. The central $8.3\times8.3\,{\rm arcmin^2}$ of this image is shown in
the top panel of Fig.\,\ref{fig:photo}. The black rectangle in the centre of the
panel represents the field of view of the NICMOS image.

\begin{figure}
  \centering
  \includegraphics[width=0.5\textwidth]{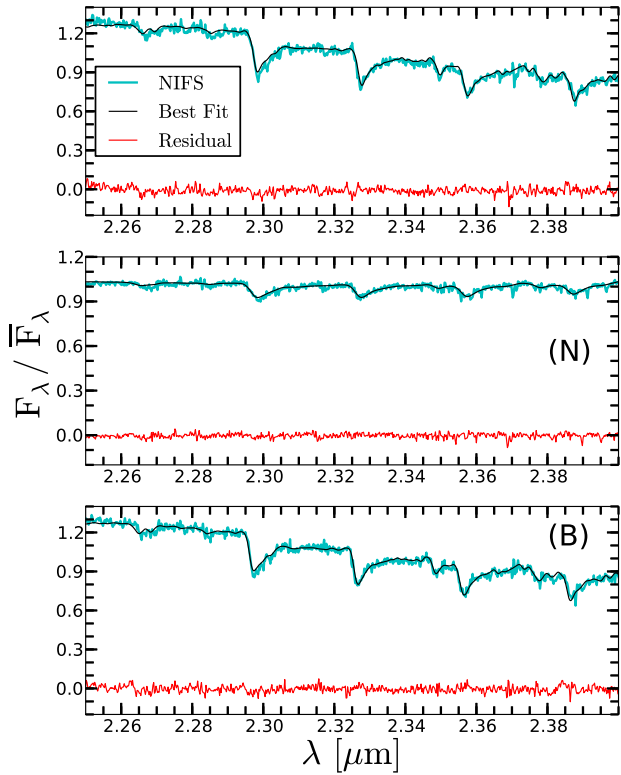}
  \caption{pPXF fits to the galaxy spectra. Examples of the resulting fit of the
pPXf to the galaxy spectra on the positions A (top panel), N (middle panel) and
B (bottom panel) indicated in Fig.\,\ref{fig:espectro}; the cyan lines are the
galaxy spectra, the black lines are the resulting fits of pPXF to the galaxy
spectra and the red lines are the residuals.}
  \label{fig:ppxf}
\end{figure}

\subsection{The NIFS data}

NIFS has a square field of view of $\approx 3.0\times3.0\,{\rm arcsec^2}$,
divided into 29 slices with an angular sampling of $0.1\,{\rm arcsec}$ in $x$
and $0.04\,{\rm arcsec}$ in $y$. The observing procedures followed the
standard Object-Sky-Object-Sky-Object-Sky-Object dither sequence, with
off-source sky positions since the target is extended, and individual exposure
times of $600\,{\rm s}$ centred at $\lambda = 2.2\,{\rm \mu m}$ with a
total of 10 individual exposures at the galaxy. The IFU was oriented with the
slices along the position angle PA$=145$\textdegree. We have used the $\rm
K_-G5605$ grating and the filter $\rm HK_-G0603$, which resulted in an arc lamp
line full width at half maximum (FWHM) of  $3.2\,{\rm \AA}$ (thus R$\sim$5300).

The data reduction was accomplished using tasks contained in the {\sc nifs}
package which is part of {\sc gemini iraf} package as well as generic {\sc
iraf} tasks.  The reduction procedure included trimming of the
images, flat-fielding, sky subtraction, wavelength and s-distortion
calibrations. We have also removed the telluric bands and flux calibrated the
frames by interpolating a black body function to the spectrum of the telluric
standard star. Small shifts between exposures due to guiding problems were
corrected by mosaicing the individual data cubes into a final one. The final
data cube contains $\sim$4300 spectra, each corresponding to an angular coverage
of $0.5\times0.5\,{\rm arcsec^2}$, which translates into
$1.75\times1.75\,{\rm pc^2}$ at the galaxy. The NIFS angular resolution with
ALTAIR in the \textit{K}-band is $0.11\,{\rm arcsec}$ as measured from the FWHM
of the spatial profile of the telluric standard stars.

In the left panel of Fig.\,\ref{fig:espectro} we present a reconstructed image 
obtained from the data cube collapsing the spectra in the same wavelength range 
of the NICMOS F222M filter, i.e. from $2.15$ to $2.28\,{\rm \mu m}$. The right
panel of the figure presents three characteristic spectra extracted from the
data cube: The nuclear spectrum (position N in the  continuum map), a spectrum
from a location at $1\,{\rm arcsec}$ NW of the nucleus (position A) and
another from $1\,{\rm arcsec}$ SE of the nucleus  (position B). The only
emission-line present in the nuclear spectrum is the broad Br$\gamma$ at $2.1661
{\rm \mu m}$. The CO absorption bands used to obtain the stellar kinematics are
identified in the spectrum from position B.

\section{Line of sight velocity distribution}\label{sec:losvd}
In order to obtain the line-of-sight velocity distribution (LOSVD) 
we have used the penalized Pixel-Fitting (pPXF) method\footnote{Available from
http://purl.org/cappellari/software} of  \cite{cappellari04} to fit the stellar
absorptions bands  present in the $K$-band spectra. The algorithm finds the
best fit to a galaxy spectrum by convolving template stellar spectra
with the corresponding LOSVD. This procedure gives as output the radial
velocity $V$, velocity dispersion $\sigma$ and higher-order Gauss-Hermite
moments h$_3$ and h$_4$.
The pPXF method allows the use of several template stellar spectra and
to vary the weights of the contribution of the different templates to
obtain the best fit, minimizing the template mismatch problem. We use the
templates from the spectroscopic library of late spectral type
stars\footnote{Available from:
http://www.gemini.edu/sciops/instruments/nearir-resources/?q=node/ 10167}
observed with the  Gemini Near Infrared Spectrograph \citep{winge09},
which have an almost identical spectral resolution to that of our
data. We have used 60 templates in the pPXF fits, and for all spaxels of
the Monte-Carlo simulations. We have also restricted the fit to the interval
shown in Fig. 4, as for smaller wavelengths there are no features to constrain
the stellar kinematics and there may be some contamination from emission lines.

\begin{figure*}
  \centering
  \includegraphics[width=1.0\textwidth]{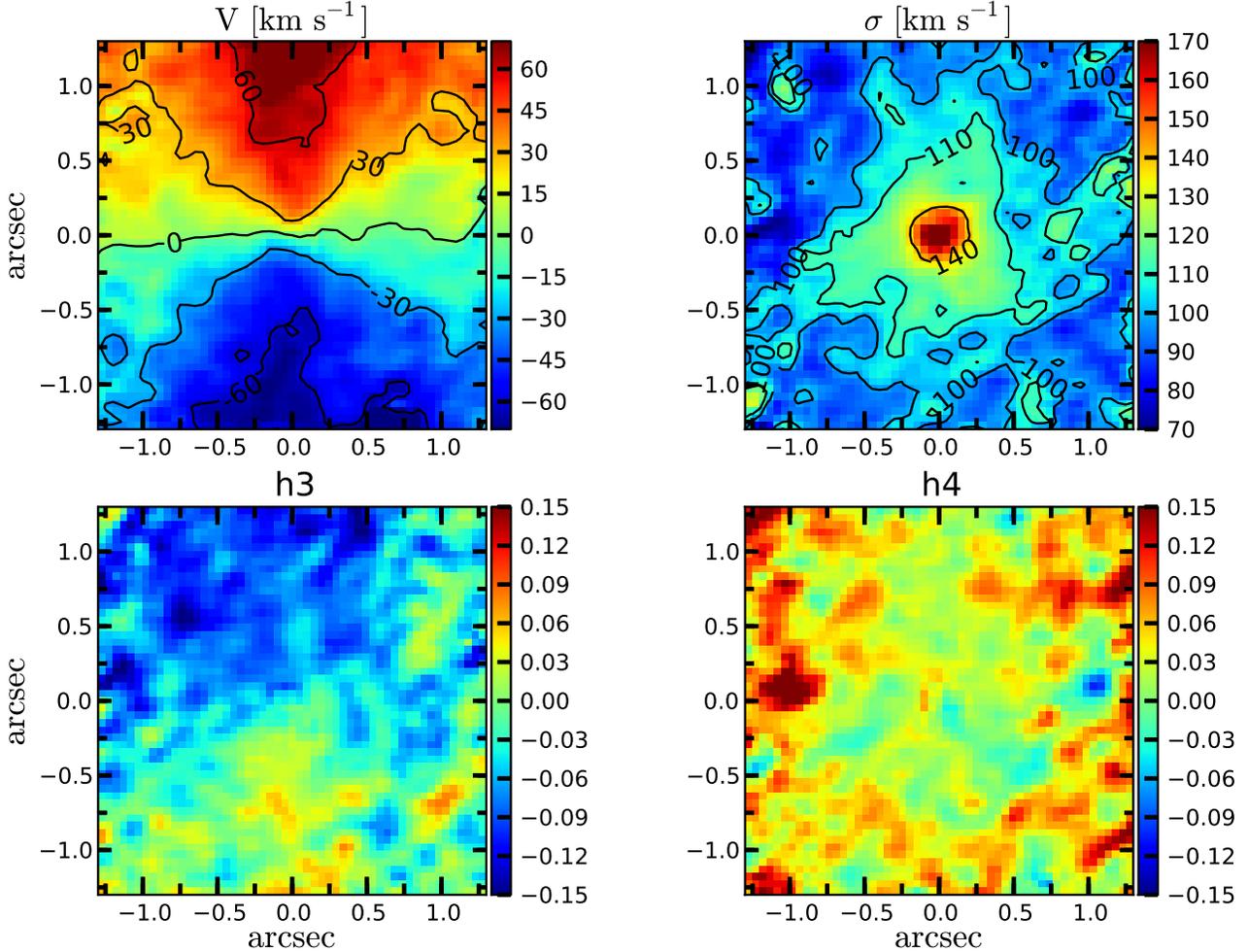}
  \caption{Two-dimensional stellar kinematic maps of NGC\,4258. Top left:
centroid velocity field. Top right: velocity dispersion. Bottom:
Gauss-Hermite moments $h_3$ (left) and $h_4$ (right). Orientation as in
Fig.\,\ref{fig:espectro} where the negative y-axis is at 145 degrees relative to
the North direction.}
  \label{fig:nifs}
\end{figure*}

\begin{figure*}
  \centering
  \includegraphics[width=1.0\textwidth]{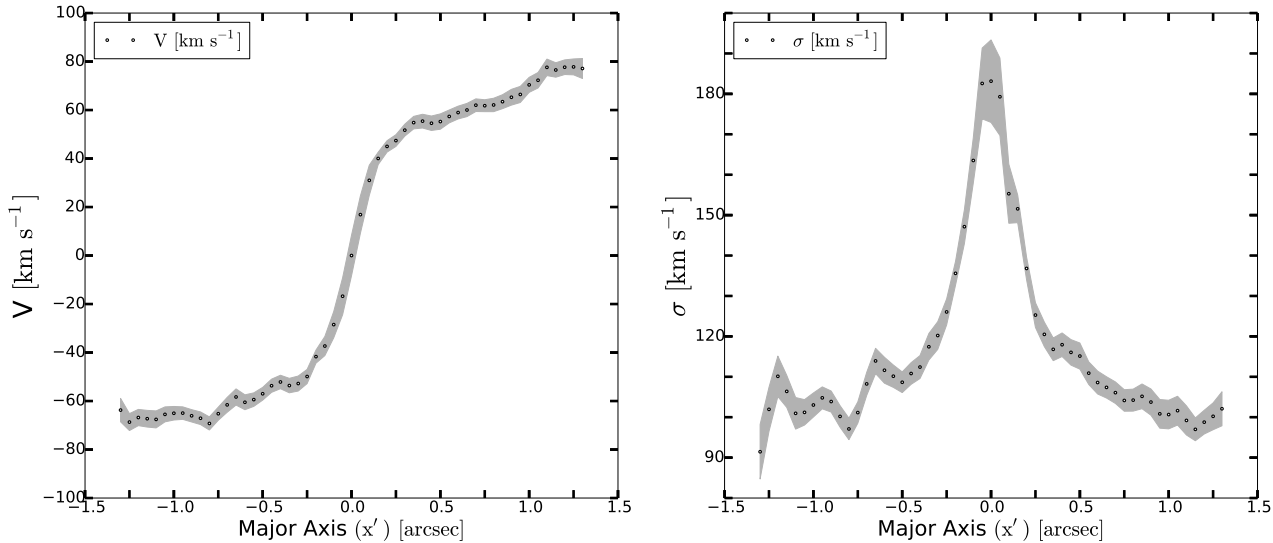}
  \caption{Measured values and Monte Carlo error estimation for the
velocities and velocity dispersions along the galaxy major axis. Left: Measured
values of the velocities $V$ with the 1$\sigma$ error bar obtained from 850
Monte Carlo simulations. Right: Measured values of the velocity dispersions
with the 1$\sigma$ error bar.}
  \label{fig:erro}
\end{figure*}

In Fig.\,\ref{fig:ppxf} we present the resulting pPXF fits to the galaxy spectra
from the positions A (top panel), N (middle panel) and B (bottom panel)
indicated in Fig.\,\ref{fig:espectro}. The cyan lines are the galaxy
spectra, the black lines are the pPXF fits to the galaxy spectra
and the red lines are the residuals. The high signal-to-noise ratio of the IFU
data, in the range of $30$ to $130$ with a average signal-to-noise ratio of
$\sim50$,  allows us to obtain fits of very high quality to
the observed galaxy spectra without the need for spatial binning. The resulting
kinematic maps of the velocity distribution are presented in
Fig.\,\ref{fig:nifs}. The top left panel shows the velocity field after
subtraction of the galaxy systemic velocity of 453\,km\,s$^{-1}$, determined as
the stellar velocity at the location of the peak of the continuum. The centroid
velocity field presents a maximum of $\sim80$\,km\,s$^{-1}$ and a rotation
pattern with the line of nodes oriented along the position angle
$\text{PA}=145$\textdegree \,and with the SE side approaching and the NW side
receding. The top right panel shows the map of the stellar velocity dispersion,
varying in the range $80-180$\,km\,s$^{-1}$, with a maximum at the 
location of the peak of the continuum. The average velocity
dispersion over the whole field of view is $\sim\,105$\,km\,s$^{-1}$. In
appendix\,\ref{app:kinematics_stis} we make a comparison of the resulting
kinematics with that presented in the paper of \cite{siopis09}, obtained
with the HST/STIS instrument. As pointed out in the Introduction,  the
radius of influence of the SMBH in NGC4258 is R$_{\text{inf}}\approx\,$15\,pc, 
that corresponds to 0.42\,arcsec at the galaxy distance. Fig.\,\ref{fig:nifs}
shows that inside this radius there is indeed a sharp rise in the $\sigma$ 
values as expected for the region  where the SMBH dominates the gravitational 
potential, what is probably better seen in Fig.\,\ref{fig:erro}.The lower panels 
show the Gauss-Hermite $h_3$ and $h_4$ moments with values ranging from $-0.15$ to
$0.15$.

The errors in the centroid velocities and velocity dispersions were
calculated from 850 Monte Carlo simulations and are illustrated in
Fig.\,\ref{fig:erro}. The left panel of this figure presents the values of the
measured centroid velocity ($V$) obtained from a cut along the galaxy major
axis (black points) and the  1$\sigma$ error bar from the Monte Carlo
simulation (shaded band). The average standard deviation of the
velocities is $3.3$\,km\,s$^{-1}$, with a maximum  of $8.4$\,km\,s$^{-1}$ in the
central pixel. Ignoring the region around the galaxy minor axis, where the
velocities are approximately zero, the average velocity error is lower than
$10\,{\rm per\,cent}$. The right panel shows de values of the measured
velocity dispersions ($\sigma$) (black points) and the 1$\sigma$ error bar
(shaded band) along the galaxy major axis. The average standard
deviation in the velocity dispersion is  $4.1$\,km\,s$^{-1}$, with a maximum of
$10.2\,{\rm per\,cent}$ in the central pixel, an average error
corresponding to $4\,{\rm per\,cent}$. The errors in the central region of
the data cube are higher than at the edges because of the presence of absorption
features stronger in the nucleus than outside, that we attribute to the effect
of dust and dilution by a red continuum from  the AGN dusty torus.

\section{Dynamical Modelling }\label{sec:model}

\subsection{Surface brightness distribution}\label{sec:mge}
In order to model the galaxy surface brightness distribution we use
the  {\sc mge$_-$fit$_-$sectors}  package\footnote{Available from
http://purl.org/cappellari/software} \citep{cappellari02} which performs a
Multi-Gaussian Expansion (MGE) parametrization for the observed surface 
brightness of the galaxy \citep{emsellem94}. In what follows, we present 
the most relevant steps in applying the MGE model to the galaxy NGC 4258 
(see \citet{cappellari08} for a complete description of the method).

The surface brightness distribution in the plane of the sky can be described
by the sum of a set of two-dimensional concentric Gaussians as
\begin{equation}\label{brilho}
 \Sigma(x',y') = \sum_{k=1}^{N}\dfrac{L_k}{2 \pi {\sigma'}_k^2 q'_k}
e^{\left[ -\frac{1}{2 {\sigma'}_k^2}\left( x'^2 +
\frac{y'^2}{q'^2_k}\right)
\right] }.
\end{equation}
where $N$ is the number of the Gaussians, each with total luminosity $L_k$, axial
ratio between $0 \leq q'_k\leq 1$ and dispersion $\sigma'_k$ along the
major axis ($x'$). In this reference system, the $x'$ axis is aligned with the
galaxy photometric major axis, derived from the photometry of the 2MASS
image, which is oriented at PA=$156$\textdegree, and $z'$ points to the line of
sight. Before comparing \eqref{brilho} with the observed surface brightness
distribution it is convolved with a MGE model for the NICMOS PSF. The NICMOS
PSF was obtained with the {\sc tiny tim} software \citep{Krist2011}. The MGE
parameters of the modelled NICMOS PSF are presented in
Tab.\,\ref{tab:gauss_ncmos} of Appendix\,\ref{app:psf_models}. In
Fig.\,\ref{fig:psf_nic} we present a comparison of the MGE model and the
 {\sc tiny tim} PSF. To examine how important are the effects of the
small differences between the MGE model and the {\sc tiny tim} PSF in the 
convolution procedure, we present in Fig.\,\ref{fig:psf_convolution} a
comparison between the surface brightness distribution convolved with the {\sc
tiny tim} PSF and the one convolved with the MGE model for the PSF. Fortunately
the two convolved surface brightness distributions are very similar. This
comparison show that the two are practically indistinguishable.
\begin{table}
\begin{center}
\caption{Gaussian parameters of the MGE model.}
\label{tab:gauss}
\begin{tabular}{@{}ccc@{}}
\hline
$I'_k\,[L_\odot \,pc^{-2}] $ &$ \sigma'_k\,[arcsec] $ & $ q'_k $ \\
\hline    
  $  247712.8  $&$ 0.063   $&$ 0.693 $\\
  $  28902.5	$&$ 0.280   $&$ 0.926 $\\
  $  9432.3	$&$ 0.501   $&$ 0.460 $\\
  $  10244.6	$&$ 0.981   $&$ 0.655 $\\
  $  3726.7	$&$ 2.677   $&$ 0.460 $\\
  $  2193.7	$&$ 3.525   $&$ 0.583 $\\
  $  607.5	$&$ 6.336   $&$ 0.460 $\\
  $  627.3	$&$ 9.022   $&$ 0.698 $\\
  $  387.3	$&$ 18.700  $&$ 0.658 $\\
  $  126.7	$&$ 41.920  $&$ 0.756 $\\
  $  181.2	$&$ 93.476  $&$ 0.450 $\\
  $  16.9	$&$ 247.722 $&$ 0.500 $\\
\hline                     
\end{tabular}
\end{center}
\begin{small}
Note: The first column
lists the surface brightnesses of each Gaussian in units L$_\odot pc^{-2}$;
the second column list the Gaussian dispersions along the major axis in
arcseconds; the third column lists the axial ratios of the Gaussians.
\end{small}
\end{table} 

In order to reproduce the surface brightness distribution of NGC\,4258 we fit
together the 2MASS and NICMOS \textit{K}-band images using the {\sc
mge$_-$fit$_-$sectors} package. NGC\,4258 has an angular size of
$18.6\times7.2\,{\rm arcmin^2}$ . The NICMOS image has an angular size of
$\sim 19.2\times 19.2\,{\rm arcsec^2}$ and is used to fit the surface brightness
distribution in the central region of the galaxy.  Due to the large 2MASS PSF
we do not use the the 2MASS image in the fit of the nuclear region ($R\leq
6.0\,{\rm arcsec}$). From the radius $6.0\,{\rm arcsec}$ to the edges of the galaxy
we fit the two images (NICMOS and 2MASS) together. As NGC\,4258 is a spiral
galaxy with a small bulge and a large disc component we implemented a
two-component MGE parametrization by separating the Gaussians in two sets with
different constraints for the axial ratios ($q'_k$). The first with $ 0.45 \leq
q'_k \leq 0.46$ to model the disc component, and the second set with more
flexible constraints, $ 0.5 \leq q'_k \leq 0.99$, to model the other galaxy
components.

\begin{figure}
  \centering
  \includegraphics[width=0.5\textwidth]{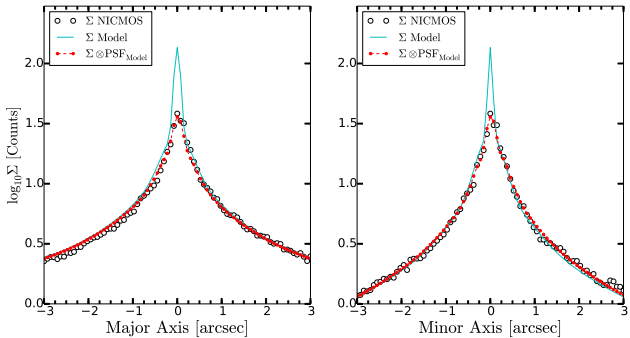}
  \caption{MGE model for the central $6.0\times6.0\,{\rm arcsec^2}$ of NGC\,4258
compared to the NICMOS image. Left: cuts along galaxy major axis of the modelled
surface brightness distribution (blue line), observed surface brightness
distribution (black open circles) and the modelled convolved surface brightness
distribution (red points). Right: same as in the left for the galaxy minor
axis.}
  \label{fig:mge_model_cuts}
\end{figure}

After several attempts we obtained a satisfying fit of the galaxy
surface brightness distribution using a set of 12 Gaussians centred in the
galaxy  nucleus and with the major axis aligned with the galaxy photometric
major axis. In Tab.\,\ref{tab:gauss} we present the parameters of the Gaussians
of our best fit: the first column lists the  surface brightnesses ($I'_k =
\frac{L_k}{2\pi\sigma'^2_k q'_k}$) in units of ${\rm L_\odot pc^{-2}}$, the
second column lists the Gaussian dispersions $\sigma'_k$ in ${\rm arcsec}$ and
the third column lists the axial ratios $q'_k$. In
Fig.\,\ref{fig:mge_model_cuts} we present linear cuts across the galaxy centre:
along the galaxy photometric major axis in the left panel and along the
photometric minor axis in the right panel. The cyan continuous lines correspond
to the MGE surface brightness distribution, the black open circles correspond to
the observed brightness distribution in the NICMOS image and the red points
correspond to the MGE after convolution with the NICMOS PSF
($\Sigma\otimes$PSF). In both panels the modelled convolved surface
brightness distribution provides a good reproduction of the observed one for the
inner 6 arcseconds of the galaxy that is approximately two times the size of the
region where we have kinematic measurements. Due the presence of the spiral arms
and of isophotal twist the surface brightness distribution in the outer region
of the galaxy can not be well reproduced by the MGE model.
The good agreement of the red points with the open circles shows the good
quality of the MGE model in describing the galaxy surface brightness
distribution.

\subsubsection{Mass density and gravitational potential }\label{subsec:rho-phi}
In order to obtain the deprojected three-dimensional luminosity density
we adopt the approximation that the galaxy is axisymmetric. In this
case the luminosity density can be obtained from the parameters that describe
the projected surface luminosity density. Assuming that the galaxy is
oblate-axisymmetric, the luminosity density can be obtained
from
\begin{equation}\label{eq:lum}
 \nu(R,z) = \sum_{k=1}^{N}\dfrac{L_k}{\left( \sqrt{2 \pi} \sigma_k\right)^3
q_k}
e^{\left[- \frac{1}{2 \sigma_k^2}\left( R^2 + \frac{z^2}{q^2_k}\right) \right]}
\end{equation}
where, $ q_k =\frac{\sqrt{q_k'^2 - \cos^2 i}}{\sin i} $, and $i$ is the galaxy
inclination ($i=90$\textdegree \,being \textit{edge-on}). It is important to
make clear that this approach can not solve the intrinsic degeneracy of the
deprojection.

The galaxy mass density can be described by a set of Gaussians as
\begin{equation}\label{eq:materia}
 \rho(R,z) = \sum_{j=1}^{M}\dfrac{M_j}{\left( \sqrt{2 \pi} \sigma_j
\right)^3 q_j} e^{\left[- \frac{1}{2 \sigma_j^2}\left( R^2 + \frac{z^2}{q^2_j}
\right) \right] }.
\end{equation}
In the self-consistent case, i. e., if only stars contribute, we can obtain
the galaxy mass density by the multiplication of the luminous density by a dynamical
mass-to-light ratio. In this case the Gaussians in \eqref{eq:materia} are the
same as in \eqref{eq:lum} with $M_j = \Gamma_k\,L_k$. The gravitational
potential generated by this density can be obtained through the Poison equation,
$\nabla^2 \Phi = 4\pi G\rho $.

The contribution of the SMBH to the gravitational potential is modelled by
the approximation that the corresponding distribution of matter is given by an
extremely narrow spherical Gaussian. The dispersion of the Gaussian is
constrained by the resolution of the kinematic data, such that $3\sigma_\bullet
\leq R_{min}$, where $R_{min}$ is the smallest distance from the black hole that
one needs to model. 

\subsection{The Jeans anisotropic dynamical model}\label{sec:jeans}
\begin{figure*}
  \centering
  \includegraphics[width=1.0\textwidth]{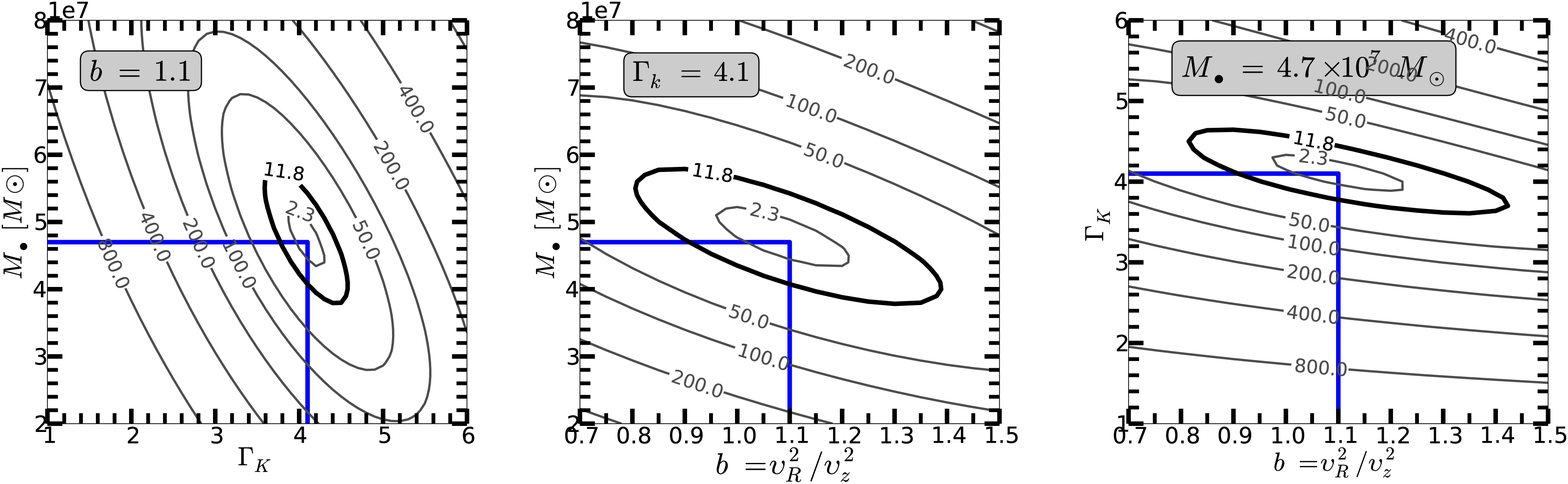}
  \caption{ Determination of the best-fitting parameters and the $\chi^2$
minimization for the model C. Left: the plane $M_\bullet\times\Gamma_K$ for
b\,=\,1.1 ($\beta_z$\,=\,0.09). Central: the plane $M_\bullet\times b$
for $\Gamma_K$\,=\,4.1. Right: the plane $\Gamma_K\times b$ for 
$M_\bullet=4.7\times10^7\,{\rm M_\odot}$. The values shown
at the contours are for the quantity $\Delta \chi^2=\chi^2-\chi^2_{min}$ where
$\chi^2_{min}$=25.48. The blue lines indicate the best-fitting parameters in
each plane.  The two inner contours shown correspond to the values of
$\Delta \chi^2$\,=\,2.3 and $\Delta \chi^2$ \,=\,11.8 that are the boundaries of
the 1$\sigma$ and 3$\sigma$ confidence regions for the two parameters
together (\citet{press}). }
  \label{fig:chi2min}
\end{figure*}

A complete description of the derivation of the Jeans equations
\citep{jeans22}, can be found in the book of  \citet{2008Binney}.
Here we only provide an overview of some fundamental assumptions of the model.
The model assumes that the galaxy is a large system of stars with positions
($x$) and velocities ($\upsilon$) described by a distribution function (DF). If
this system is in a  steady state under the influence of a smooth
gravitational potential, the DF must satisfy the steady-state collisionless
Boltzmann equation.  Under the assumption that the galaxy has axial symmetry
we obtain the axisymmetric Jeans equations in cylindrical coordinates
(\cite{2008Binney}, equations 4-29). The solutions of these equations provide
the velocity moments of the DF and can be compared with the observed kinematics
of the galaxy.

Semi-isotropic Jeans dynamical models \citep{jeans22} of galaxies have been
applied  for different purposes \citep{Nagai1976, Satoh1980, 
Binney1990,  Marel1990, emsellem94, Albada1995, Riciputi2005}. Two of the most
interesting applications are to estimate the dynamical mass-to-light ratio of
galaxies \citep{Marel1991, Statler1999, Cappellari2006, Cortes2008} and to
obtain the masses of the SMBH in the nuclei of galaxies \citep{Magorrian1998,
Marel1998, Cretton1999, Joseph2001}. 

\citet{cappellari08} has presented an anisotropic generalization of the
axisymmetric semi-isotropic Jeans formalism. Making the assumption that the
velocity ellipsoid is aligned with a cylindrical coordinate system ($R,\, z,\,
\phi$) and that the anisotropy is constant and quantified
by $\overline{\upsilon_R^2}=b\,\overline{\upsilon_z^2}$, the Jeans
equations reduce to (eqs. (8) and (9) of \cite{cappellari08}):
\begin{equation}
 \dfrac{b\, \nu \overline{\upsilon_z^2} - \nu \overline{\upsilon_{\phi}^2}}{R} +
\dfrac{\partial (b\, \nu \overline{\upsilon_z^2} )}{\partial R} = -\nu
\dfrac{\partial \Phi}{\partial R}
\end{equation}
\begin{equation}
 \dfrac{(\nu \overline{\upsilon_z^2})}{\partial z} = -\nu\dfrac{\partial
\Phi}{\partial z}
\end{equation}

One advantage of this approach is that, if a MGE model for the galaxy surface
brightness is available, the solutions of the Jeans equations, (i. e., the first
and second moments in velocity of the distribution function (DF) of the system,
and the projections of these  moments in the plane of the sky) are
given as a function of the Gaussian parameters of the luminosity density
\eqref{eq:lum} and mass density \eqref{eq:materia}. \citet{cappellari08} applied
this method to determine the mass-to-light ratio and inclination of galaxies
classified as fast-rotators in the SAURON survey \citep{deZeeuw2002}. The
effectiveness of this Jeans Anisotropic Multi-Gaussian expansion dynamical model
(JAM)\footnote{Available from http://purl.org/cappellari/software} is shown in
\citet{Cappellari2009} where they model the observed
kinematics of Centaurus\,A with the same parameters ($M/L, M_\bullet$)
obtained from the Schwarzschild orbit superposition method. Another test of
the JAM models is presented by \citet{Lablanche2012} where the authors
demonstrate the ability of the models in reproducing the anisotropy profile and
mass-to-light ratios of realistic N-body collisionless simulations of barred and
unbarred galaxies. In \citet{Medling2011} the JAM models are used to determine
the mass of the  SMBH in he southern component of the pair of interacting
galaxies in NGC\,6240. In \citet{Neumayer2012} they established upper limits for
the black hole masses of 9 late type galaxies using the JAM models.
\citet{Raimundo2013} uses the JAM models to obtain an upper limit of the black
hole mass in MCG-6-30-15. \cite{Emsellem2013} used the JAM models together with
N-body simulations to review the hypothesis of the presence of an overmassive
black hole in NGC\,1277. Another frequent application of the JAM models is the
determination of intermediate-mass black holes in globular clusters 
\citep{Seth2010, Lutzgendorf2011, Lutzgendorf2012a, Lutzgendorf2013,
Feldmeier2013}. The largest applications of the JAM method to
date are the studies of the stellar initial mass function
\citep{Cappellari2012Natur} and dynamical scaling relations
\citep{Cappellari2013MNRAS4321862C} of the 260 early-type galaxies of the
ATLAS$^{\rm 3D}$ survey.

The anisotropic Jeans method used in this work is significantly
different from Schwarzschild method. The former describes the orbital
distribution via a few anisotropy parameters, while the latter makes virtually
no assumptions on the orbital distribution. A drawback of the Jeans approach is
that it can potentially bias the BH mass determination. However this issue is
largely overcome by allowing for a variation in the ansiotropy, which can be
constrained by integral-field kinematics, as we do here. Previous tests agree
with the result of this paper, showing a general consistency between the
Schwarzschild and JAM approach \cite{Cappellari2010, Seth2014}. An advantage of
the Jeans approach is its good predictive power: It is unable to fit the noise
of the systematics in the data and can often be used to detect problems with the
data or flag bad data (e.g. \citet{2013MNRAS.432.1709C}). The Jeans solutions
are also relatively easy to qualitatively understand. The Schwarzschild
approach, given its generality in the adopted orbital distribution, does not
suffer from potential bias in the BH mass. However the method can easily fit
noise and bad data without raising any concern. The method can easily create
unphysical orbital distributions outside the region constrained by the
kinematics (e.g. Fig.\,2 of \citet{2005CQGra..22S.347C}). For this reason
it provides better constraints to the SMBH masses  when integral-field data are
available out to larger radii \citep{2002MNRAS.335..517V}. Overall, the two
methods are complementary  and sufficiently different to motivate the
re-determination of BH mass presented in this paper, using Jean rather than the
previously published Schwarzschild approach.

\subsubsection{The velocity second moment}
For a galaxy with a surface brightness distribution parametrized by a MGE model
as in sec.\,\ref{sec:mge}, the projection in the plane of the sky of the
second moment, $\langle \upsilon_{los}^2  (x',y') \rangle$  of the DF is
given in terms of the parameters of the Gaussians (eq. (28) of
\cite{cappellari08}). This quantity is a function of three free parameters: the
galaxy mass-to-light ratio ($\Gamma_K$), the anisotropy ($\beta_z =
1-\frac{\overline{\upsilon_z^2}}{\overline{\upsilon_R^2}}$) and the mass of the
SMBH ($M_\bullet$). The comparison of the modelled $\langle \upsilon^2_{los}
\rangle^{1/2}$ with the measured $V_{rms} = \sqrt{V^2 +\sigma^2}$, where $V$ and
$\sigma$ are shown in Fig.\,\ref{fig:nifs}, provide the values of the
best-fitting parameters for the galaxy.

We start by assuming that the galaxy has a constant
anisotropy in the velocity dispersions. Thus the space of parameters has three
dimensions: $M_\bullet \times \Gamma_K \times \beta_z$. We then perform a
$\chi^2$ minimization in this space searching for the values of
$M_\bullet,\,\Gamma_K$ and $\beta_z$ that best fit the observed kinematics. We
weight the $\chi^2$ minimization by assigning errors to the kinematic
measurements that are inversely proportional to the galaxy surface brightness
at each location. This is done in an attempt to give comparable
contributions to the $\chi^2$ for all radii sampled by the NIFS kinematics.
Without this weighting, the $\chi^2$ would be artificially dominated by the
numerous pixels at large radii, which contain virtually no information on the
mass of the central supermassive black hole. In our trials without
weighting the kinematics in the minimization we have obtained a significantly
larger value for the mass of the SMBH ($M_\bullet=7.2\times10^7\,{\rm
M_\odot}$), but the modelled values of the velocity second moment clearly did
not reproduce the values of the observed ${\rm V_{rms}}$ for the central region
of the galaxy.

We considered in a first trial that the galaxy has an  inclination of
$i=72$\textdegree \, (Model A). This inclination is obtained considering that
the outer disc of the galaxy ($r=200\,{\rm arcsec}$) is thin \citep{vanAlbada80,
siopis09}. The best-fitting model for this inclination with
$\chi_{min}^2$\,=\,27.21  is obtained with the parameters
$M_\bullet=5.2\times10^7\,{\rm M_\odot}$, $\Gamma_K$\,=\,4.3 and
$ b=0.90\, (\beta_z$\,=\,-0.11). This model does not reproduce
satisfactorily the measured kinematics along the galaxy minor axis. 

After several attempts with different values for the galaxy inclination
we obtained a satisfactory model for $i=64$\textdegree \, (Model C) . The
minimum  $\chi^2$ obtained was $\chi_{min}^2$\,=\,25.48 for 
$M_\bullet=4.7\times10^7\,{\rm M_\odot}$, $\Gamma_K$\,=\,4.1 and
$b=1.1\, (\beta_z$\,=\,0.09). In Fig.\,\ref{fig:chi2min} we show the
$\chi^2$ minimization in the space of parameters. In Tab.\,2 we
list the best-fitting parameters for three models with different
inclinations: the models A and C described before and a third model
with an inclination of $i=68$\textdegree \, for which the best-fitting
parameters are $M_\bullet=5.0\times10^7\,{\rm M_\odot}$, $\Gamma_K$\,=\,4.3 and
$b=0.95\, (\beta_z$\,=\,-0.05), that provide a minimum $\chi^2$ of 26.84.
Systematic variations in the best-fitting parameters are observed as the  galaxy
inclination decreases: the values of the mass of the SMBH and the mass-to-light
ratio decrease but the value of the anisotropy parameters increase.

\begin{table}
\label{tab:models_const}
\caption{Summary of the models with constant anisotropy.}
\begin{center}
\begin{tabular}{@{}cccccc@{}}
\hline
Model & $i$ & $M_\bullet$ [$M_\odot$]& $\Gamma_K$  &  b
($\beta_z$) & $\chi^2_{min}$    \\
(1)  &  (2)  & (3)  &  (4)  & (5)  &  (6)    \\
\hline
A  &  72\textdegree& 5.2$\times10^7$ & 4.3 & 0.90 (-0.11) & 27.21  \\
B  &  68\textdegree& 5.0$\times10^7$ & 4.3 & 0.95 (-0.05) & 26.84  \\
C  &  64\textdegree& 4.7$\times10^7$ & 4.1 & 1.1 (0.09) & 25.48  \\
\hline
\end{tabular}
\end{center}
\begin{small}
 Note. Column (1): model designation. Column (2): galaxy inclination. Column
(3): best-fitting mass of the SMBH. Column (4): best-fitting \textit{K}-band
mass-to-light ratio.  Column (5): best-fitting anisotropy. Column (6): lowest
value of the $\chi^2$ obtained.
\end{small}
\end{table}

\begin{figure}
  \centering
  \includegraphics[width=0.5\textwidth]{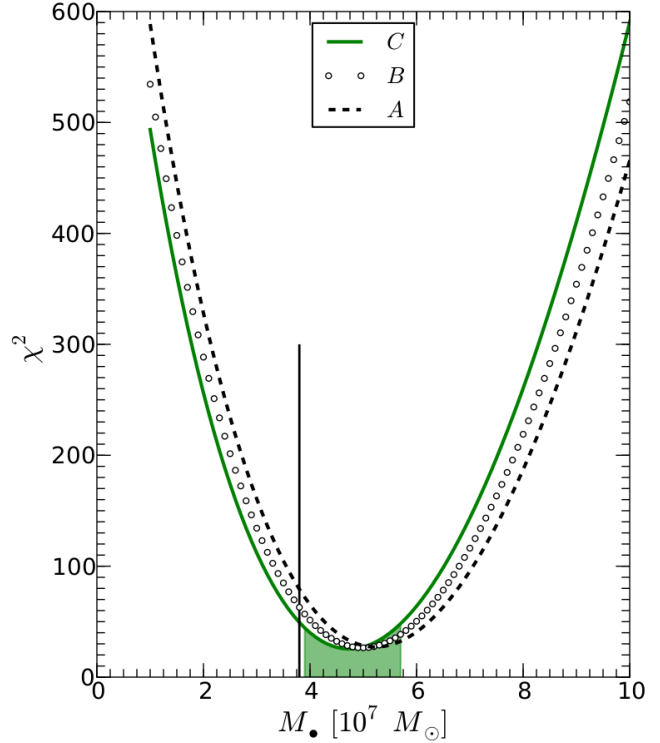}
\caption{  Variation of $\chi^2$ with the SMBH mass for the models with
constant anisotropy. The green solid line represents the minimization of
$\chi^2$ as a function of $M_\bullet$ for the best-fitting model with
$i=64$\textdegree \,(Model C). The values of $M_\bullet$ in the shaded region
correspond to the 3$\sigma$ uncertainty in this parameter, thus the
best-fitting mass of the SMBH is $M_\bullet =
4.7^{+1.0}_{-0.8}\times10^7\,\,{\rm M_\odot}$.
The open circles represent the minimization for the model with
$i=68$\textdegree \,(Model B). The black dashed line represents the 
minimization for the model with $i=72$\textdegree \, (Model A). The vertical
line indicates the value of the maser determination, $M_{\bullet\,{\rm Maser}} =
3.82\times10^{7}\,\,{\rm M_\odot}$.}
\label{fig:best_incr_lin}
\end{figure}

In Fig.\,\ref{fig:best_incr_lin} we present the $\chi^2$ minimization for
$M_\bullet$  for models A (black dotted line), B (open circles) and  C
(green solid line). The values of the parameters $\Gamma_K$ and $b$ are the best-fitting values of
each model. The vertical line indicates the value of the maser determination,
$M_{\bullet\,{\rm Maser}} = 3.82\times10^{7}\,\,{\rm M_\odot}$. The green shaded
region corresponds of the $M_\bullet$ values that are within 3$\sigma$
confidence intervals for the model with $i=64$\textdegree. The best-fitting mass
of the SMBH with 3$\sigma$ of confidence is $M_\bullet =
4.7^{+1.0}_{-0.8}\times10^7\,\,{\rm M_\odot}$.

\subsubsection{Effects of the PSF in the best-fitting parameters}

One important issue in the dynamical models of galaxies is that before comparing
the modelled velocities with the measured ones we need to convolve the
modelled velocities with the PSF of the kinematic observations. For the Jeans
anisotropic dynamical models the convolution is implemented as in Appendix A
of \cite{cappellari08}. We used a model for the NIFS PSF that is obtained from
a MGE fit to the observations of three stars observed in the same night of the
galaxy observations, as described in sec. \ref{app:psf_nifs}. As the
time needed for the galaxy observations are longer (600\,s) than that of
the stars observations (15\,s) there is the possibility that we are
underestimating the NIFS PSF.

In order to verify the influence of a possible broader PSF (larger FWHM) in
the modelled kinematics and best-fitting value for $M_\bullet$ we ran again the
model C with  $\Gamma_K=4.1$, and  $b=1.10$ and performed the $\chi^2$
minimization for $M_\bullet$. We used two broader PSF models obtained 
from the model described in sec.\,\ref{app:psf_nifs}: in the first one we
increase the dispersions (listed in second column of Tab.\,\ref{tab:gauss_nifs})
of the Gaussians of the MGE model of the PSF by a factor of 1.5  and in the
second we duplicated the values of the dispersions of the Gaussians.
\begin{figure}
  \centering
  \includegraphics[width=0.5\textwidth]{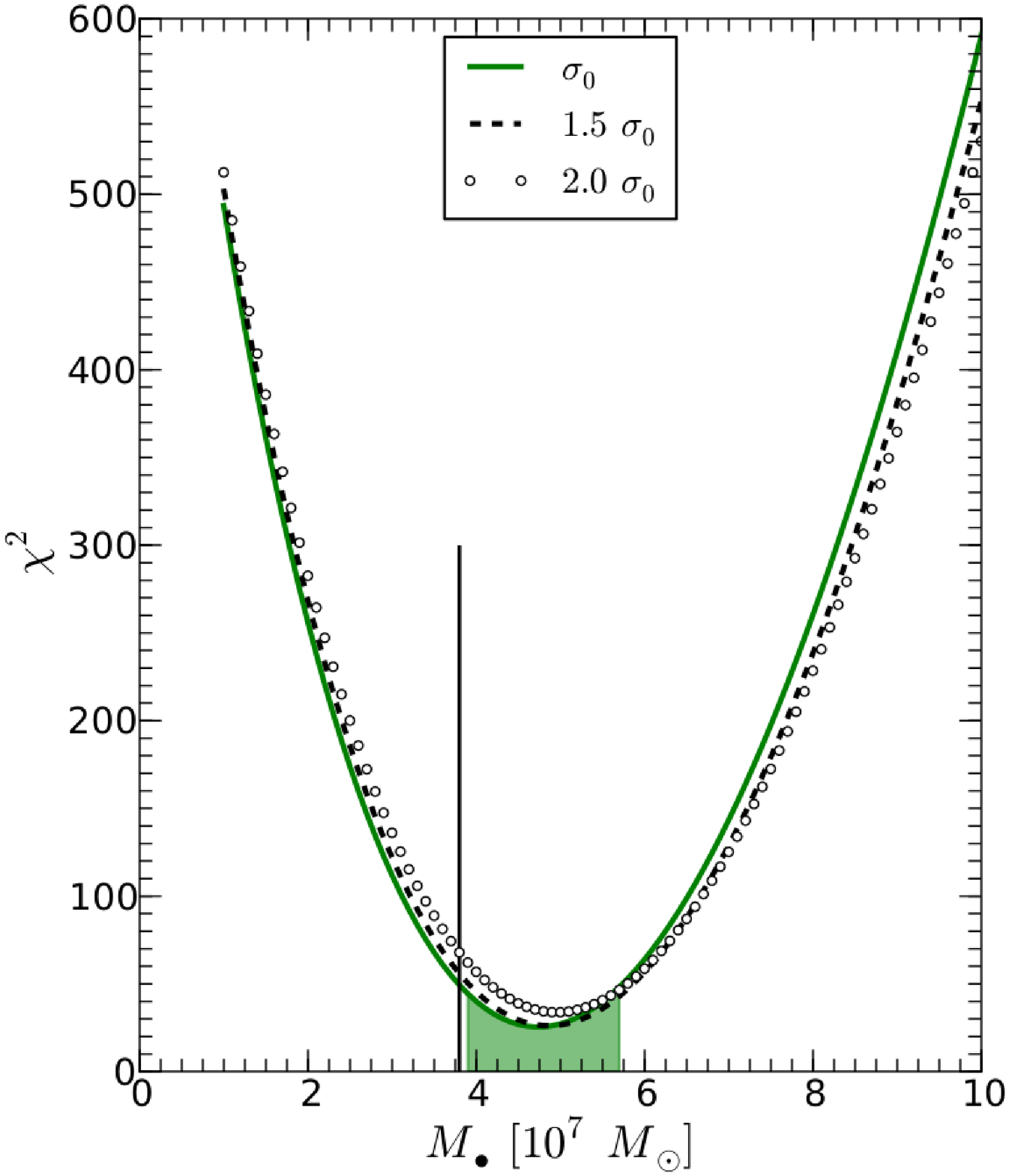}
\caption{ Effects of variations in the PSF width in the best-fitting mass of the
SMBH. The green solid line represents the variation of $\chi^2$ with the mass of
the SMBH for the model C. The dashed black line shows the variation of $\chi^2$
using a model for the NIFS PSF with a dispersion $\sigma$ that is $50\,{\rm
per\,cent}$ larger than that presented in App.\,\ref{app:psf_nifs}, which
results in a $\chi^2_{min}=26.17$ for $M_\bullet =4.9\times10^7\,{\rm M_\odot}$.
The open circles shows the variation of $\chi^2$ using a model for the NIFS
PSF with a dispersion $\sigma$ that is $100\,{\rm per\,cent}$ larger than the
original and results in a $\chi^2_{min}=33.83$ for $M_\bullet
=5.0\times10^7\,{\rm M_\odot}$. The shaded region shows the range of values of
$M_\bullet$ that are inside the 3\,$\sigma$ uncertainty of the best-fitting
value of $M_\bullet$ of the reference model (Model C). The vertical line
indicates the value of the maser determination, $M_{\bullet\,{\rm Maser}} =
3.82\times10^{7}\,\,{\rm M_\odot}$. }
\label{fig:best_lin_psf}
\end{figure}

In Fig.\,\ref{fig:best_lin_psf} we present the modifications in the
$\chi^2$ minimization and in the best-fitting value of $M_\bullet$ introduced
by the increase of the NIFS PSF by a factor of 1.5 (represented by the black
dotted line) and by a factor of 2.0 (represented by the open circles). The
continuous line  is for model C which provides a $\chi^2_{min} = 25.48 $ for
$M_\bullet= 4.7\times10^7\,{\rm M_\odot}$. For the model with the PSF 1.5 times
broader we obtained $\chi^2_{min} = 26.17$  for $M_\bullet =4.9\times10^7\,{\rm
M_\odot}$ which is within the 1$\sigma$ uncertainties of the best-fitting
model.  For the model with the PSF 2 times broader $\chi^2_{min} = 33.83$
for $M_\bullet =5.0\times10^7\,{\rm M_\odot}$. For both cases the values for
the mass of the SMBH that best reproduce the measured kinematics are inside the 
3$\sigma$ confidence interval of the best-fitting model and the modelled 
velocities still reproduce the observed ones.

\subsubsection{Models with variable anisotropy}
\begin{figure*}
  \centering
  \includegraphics[width=1.0\textwidth]{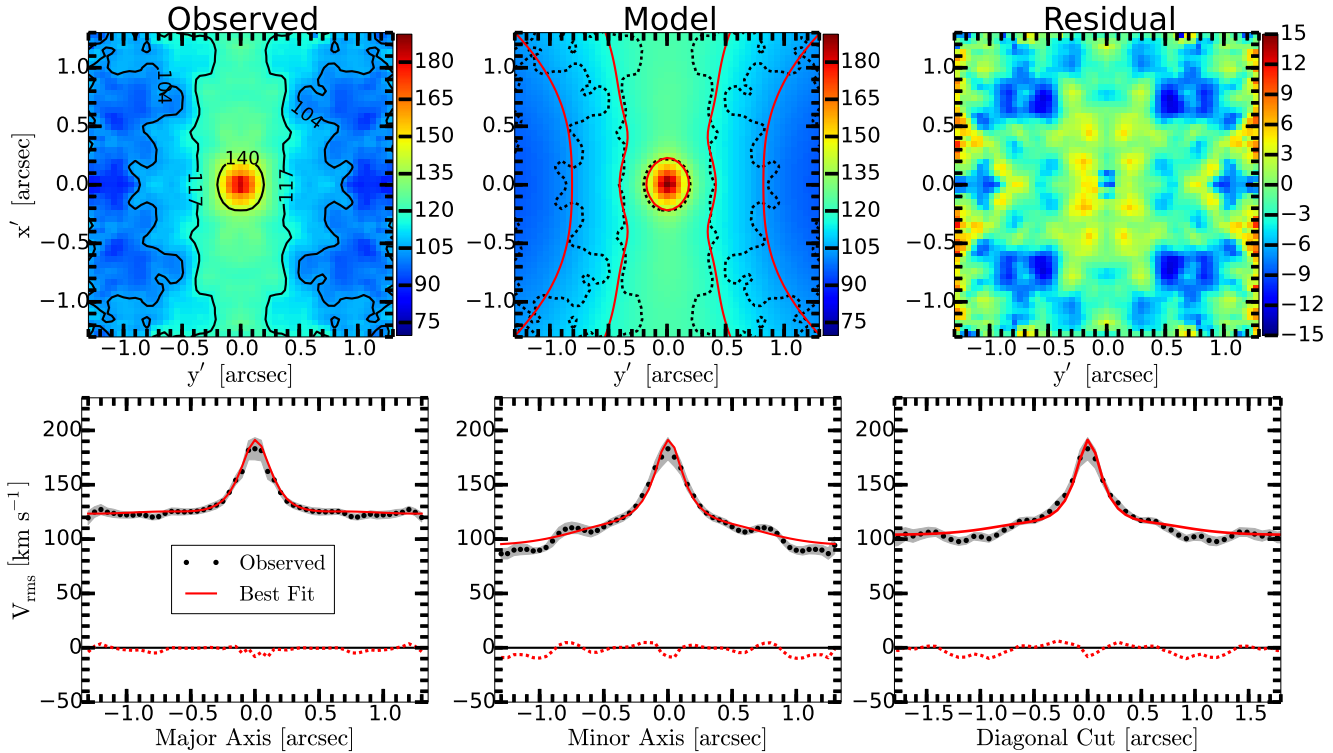}
\caption{Comparison between the observed $V_{rms}$ and the best-fitting
$\langle \upsilon^2_{los} \rangle^{1/2}$ (Model D). Top left: the observed
$V_{rms}=\sqrt{V^2 +\sigma^2}$ after symmetrization; the contours
correspond to the  isovelocity curves of $104, 117$ and $140\,{\rm km s^{-1}}$.
Top central: the best-fitting model presenting the superposition of the
isovelocity curves of the modelled (red solid lines) and observed (black dotted
lines) velocities. Top right: the residual $V_{rms} - \langle \upsilon^2_{los}
\rangle^{1/2}$ in ${\rm km s^{-1}}$. Bottom left: linear cuts along the galaxy
major axis ($x'$), where the black points show the observed symmetrized
$V_{rms}$, the red solid lines are the result of our best-fitting model, the
shaded bands are the 1$\sigma$ errors in the velocity measurements. Bottom
central and right panels show the same quantities as the left panel along the
galaxy minor axis and along the diagonal direction, respectively.}
\label{fig:best-con}
\end{figure*}
 As the solutions of the Jeans equations are presented for each Gaussian
individually and the total second moment is the quadratic sum of the
contributions of each Gaussian, it is possible to assign different values of
anisotropy for each Gaussian of the MGE model that describes the galaxy
luminosity density. As exemplified by \citet{krajnovic2009} for the
galaxies NGC\,2549 and NGC\,524 and by \citet{2003ApJ...583...92G}, spatial
variations in the velocity anisotropy are frequent. In order to test how better
the measured kinematics can be 
reproduced by models with variable anisotropy, we considered two scenarios. In
the first we assumed that there is a radial variation in the anisotropy. We
considered two different  values for the anisotropy parameters
$b_k$, one for the four inner Gaussians that have $\sigma_k < 1.0\,{\rm
arcsec}$ and another value for the remaining outer Gaussians. Then we performed
the $\chi^2$ minimization  for two different inclinations $i=64$\textdegree \,
(model D) and $i=72$\textdegree \,(model E). In a second scenario we assigned
one anisotropy value for the Gaussians with axial ratio $q_k < 0.5$
representative of the disc component and another anisotropy value for the
Gaussians with axial ratio $q_k \geq 0.5$ representative of the spheroidal
component. This minimization was also performed for two values of the galaxy
inclination: $i=64$\textdegree \,(model F) and $i=72$\textdegree \,(model G).
For all the models with variable anisotropy we considered a constant
mass-to-light ratio of $\Gamma_k=4.1$. The best-fitting parameters for the
models with variable anisotropy are presented in Tab.\,\ref{tab:models_var}.
\begin{table}
\caption{Summary of the models with variable anisotropy}
\label{tab:models_var}
\begin{tabular}{@{}ccccccc@{}}
\hline
Model & $i$ & $M_\bullet$ [${\rm M_\odot}$]& b$_{in}$ ($\beta_{z}$) & b$_{out}$
($\beta_{z}$) & $\chi^2_{min}$     \\
(1)  &  (2)  & (3)  &  (4)  & (5)  &  (6)     \\
\hline
D  &  64\textdegree& 4.8$\times10^7$ & 1.10 (0.09)& 1.05 (0.05) & 25.24 \\
E  &  72\textdegree& 5.3$\times10^7$ & 1.00 (0.00)& 0.90 (-0.11) & 28.65  \\
\hline
 &  & & b$_{disc}$ ($\beta_{z}$) & b$_{bulge}$
($\beta_{z}$) &      \\
\hline
F   &  64\textdegree& 4.8$\times10^7$ &  1.25 (0.20)&  1.05 (0.05) & 25.34 \\
G  &  72\textdegree& 5.3$\times10^7$  &  1.30 (0.23)&  0.85 (-0.17) & 27.90 \\
\hline
\end{tabular}
\begin{small}
 Note. Column (1): model designation. Column (2): galaxy inclination. Column
(3): best-fitting mass of the SMBH. Column (4): for models (D) and (E)
these are these are the values of the anisotropy parameter of the inner 4
gaussians, for models (F) and (G) these are the values of the anisotropy
parameter of the Gaussians with axial ratio $q_k$ lower than 0.47. Column (5):
for models (D) and (E) these are the values of the anisotropy parameter of the
outer 8 Gaussians, for models (F) and (G) these are the values of the
anisotropy parameter of the Gaussians with axial ratio $q_k$ higher than 0.5.
Column (6): lowest value of the $\chi^2$ obtained. 
\end{small}
\end{table}
The best-fitting model with  $\chi^2_{min} = 25.24$ was obtained for the 
inclination $i=64$\textdegree\, and with a radially variable anisotropy
(Model D). The dynamical parameters of this model are $M_\bullet=
4.8\times10^7\,{\rm M_\odot}$ , $\Gamma_K = 4.1$, $b_{in}= 1.10 $ and $b_{out}=
1.05$. The average absolute error in the second moment over all pixels of the
kinematic field  is of $5.8\,{\rm per\,cent}$. The results of the model for the
second moment and the comparison with the observed $V_{rms}$ are shown in
Fig.\,\ref{fig:best-con}.

\begin{figure}
  \centering
  \includegraphics[width=0.5\textwidth]{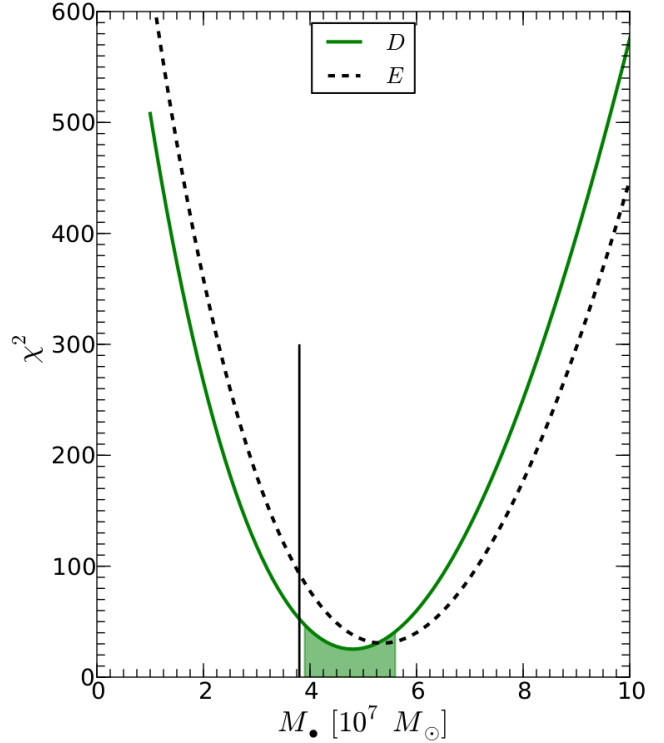}
\caption{ Variation of the $\chi^2$ with the SMBH mass for the models with
variable anisotropy. The green solid line represents the minimization of
$\chi^2$ as a function of the $M_\bullet$ for the best-fitting model with
$i=64$\textdegree \,(Model D). The values of $M_\bullet$ in the shaded region
corresponds to the 3$\sigma$ confidence interval for this parameter, thus the
best-fitting mass of the SMBH is $M_\bullet = 
4.8^{+0.8}_{-0.9}\times10^{7}\,{\rm M_\odot}$. The black dashed line is the
$\chi^2$ minimization for the model with $i=72$\textdegree \,(Model E).}
\label{fig:best_bvar_lin}
\end{figure}

The top left panel presents the observed $V_{rms}$ after bi-symmetrization. The
minimum value for $V_{rms}$ is $\approx 85\,{\rm km\,s^{-1}}$ and occurs at the
most distant locations from the galaxy centre along the photometric minor axis.
The maximum value is $\approx 180\,{\rm km\,s^{-1}}$ at the galaxy centre. 
The top central panel shows the $\langle \upsilon^2_{los}
\rangle^{1/2}$ distribution  for the best-fitting
model (model D), showing very good agreement with the measured $V_{rms}$. The
top right panel shows the residuals
$\left(V_{rms} -\langle \upsilon^2_{los} \rangle^{1/2} \right)$. The differences
between the measured and modelled velocities are small, with the highest
positive
residual being $\approx 12\,{\rm km\,s^{-1}}$ (which represents a error
of $\approx 11\,{\rm per\,cent}$ in relation to the measured velocity in this
position), and highest negative residual of $\approx-13\,{\rm km\,s^{-1}}$
representing an error of $\approx12\,{\rm per\,cent}$.  The average error over
the whole field is  $\approx3\,{\rm per\,cent}$. The good  agreement between
the modelled and measured velocities is also clear in the superposition of the
isovelocity curves of the model on those of the  measured velocity field
presented in the top central panel.

In the bottom panels we present linear cuts across the galaxy centre in three
different directions: along the galaxy major axis in the left panel, the minor
axis in the central panel and along a diagonal cut in the right panel. The
black points show the measured values for $V_{rms}$ with  1$\sigma$ error bars
represented by the shaded region. The red solid lines show our best-fitting
model for the projected second moment of the velocity
($\langle \upsilon^2_{los} \rangle^{1/2}$). The red dotted lines
show the residuals between the observations and the best-fitting
model. Along the galaxy major axis the modelled velocities reproduced very well
the observed ones. Along the galaxy minor axis the largest differences occur
in the most distant regions from the galaxy centre. 

In Fig.\,\ref{fig:best_bvar_lin} we present the $\chi^2$ minimization for 
$M_\bullet$ for the models D (green solid line) and E (black dashed line). The
vertical line indicates the value of the maser determination,
$M_{\bullet\,{\rm Maser}} = 3.82\times10^{7}\,{\rm M_\odot}$. The green shaded
region corresponds to the $M_\bullet$ values that are inside the 3$\sigma$
confidence interval for the model with $i=64$\textdegree \,(Model D). The
best-fitting mass of the SMBH with 1$\sigma$ of confidence is $M_\bullet
= 4.8^{+0.8}_{-0.9}\times10^{7}\,{\rm M_\odot}$.

\subsubsection{Comparison with previous results}

We can now compare the value of the SMBH mass that we have obtained with
the previous ones from the literature. When compared with the maser
determination \citep{greenhill95, herrnstein99} of $M_{\bullet\,{\rm Maser}} =
3.82\times 10^7\,{\rm M_\odot}$ our value of  $M_\bullet =
4.8^{+0.8}_{-0.9}\times10^{7}\,{\rm M_\odot}$ is $\approx$ 25\,{\rm
per\,cent} larger.

For comparison, the  previous stellar dynamical determination obtained
via Schwarzschild models from long-slit data by \citet{siopis09}
($M_{\bullet\,{\rm Schw}} = 3.3\pm0.2 \times 10^7\,{\rm M_\odot}$)  is 15 per
cent lower than the maser value. Making a direct comparison between the two
dynamical determinations, there is a difference of $\approx 45\,{\rm
per\,cent}$ in the best-fitting values for the mass of the SMBH. The main
factors that contribute to this difference are:

\begin{itemize}
 \item As we have shown in Fig.\,\ref{fig:kin1}
of appendix\,\ref{app:kinematics_stis} the measured velocity dispersions of our
NIFS data are considerably higher than those tabulated for the STIS data,
being the differences larger than the error bars for most of the points
analysed. The differences between the values of the V$_{rms}$ are consistent
with the differences introduced in the modelled velocity second moment by the
difference of the mass of the SMBH from the models.

\item Another possible cause for these difference in the mass of the
SMBH can be an over-subtraction of the contribution from the AGN to the surface
brightness distribution in the  NICMOS image.  The uncertainties introduced in
the model by this factor are discussed in the
appendix\,\ref{app:oversubtraction}.

\item A third factor that could cause the difference in the SMBH mass
derived using the two methods are the intrinsic differences between the
dynamical models used (Schwarszchild vs. Jeans models). The adopted JAM method
makes more restrictive assumptions than Schwarszchild method. Alought the two
methods have been shown to generally agree quite well in real galaxies, some
differences are not unexpected.

\end{itemize}

\subsubsection{The velocity first moment}

In order to model the velocity  first moment it is necessary to make extra
assumptions in the Jeans equations to specify how the second moment
is composed in terms of ordered and random motions. We use the approximation
that the tangential component of the velocity first moment of each
Gaussian is a function of the difference between the tangential and radial
components of the velocity second moment of each Gaussian, $
\langle \upsilon_\phi \rangle_k = \kappa_k \left(\langle \upsilon_\phi^2
\rangle_k - \langle \upsilon_R^2\rangle_k \right)^{1/2}$. (See Sec.\,3.2.1 of
\citet{cappellari08} for a more complete explanation). 

\begin{figure*}
  \centering
\includegraphics[width=1.0\textwidth]{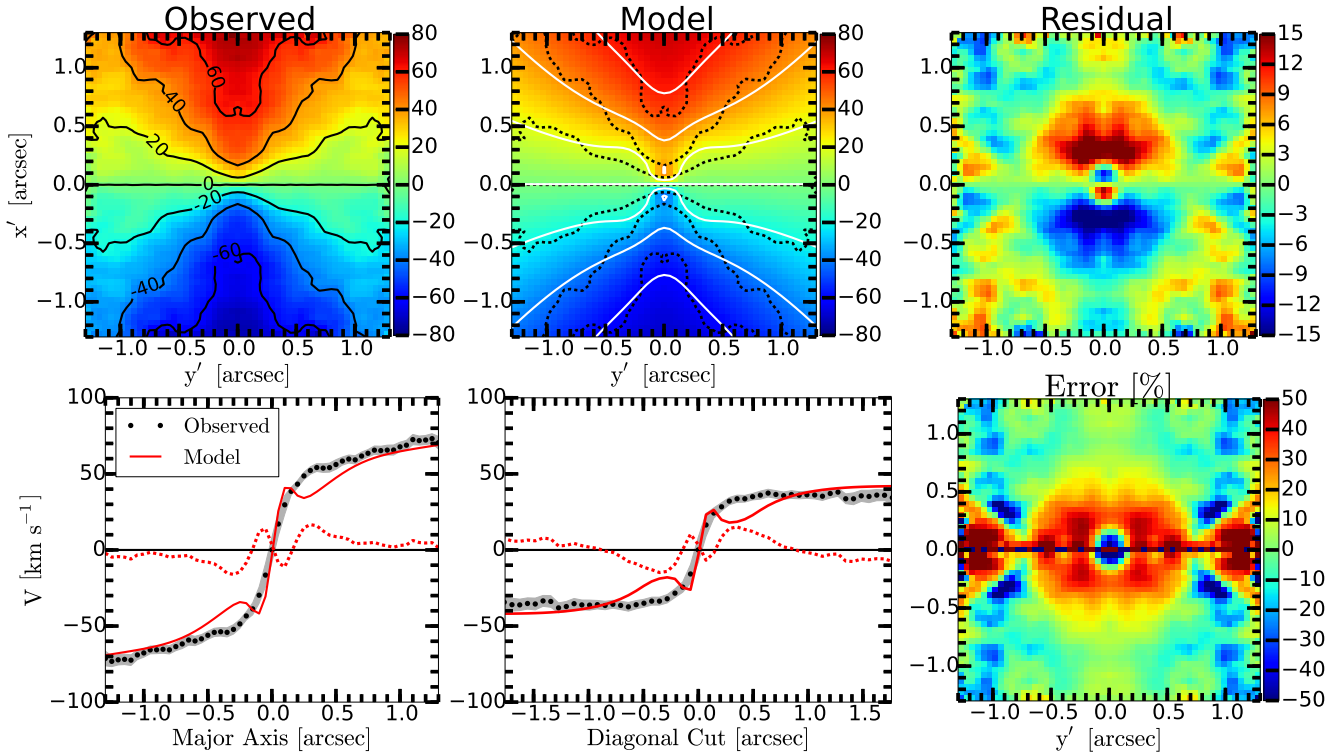}
\caption{Comparison between the measured centroid velocity V and the first
moment $\langle \upsilon_{los} \rangle$ for model D. Top left: measured velocity
$V$; the contours are for the isovelocity curves of $0.0, \pm20, \pm40,
\pm60\,{\rm km\,s^{-1}}$. Top central: modelled $\langle \upsilon_{los} \rangle$
with the best-fitting values of $M_\bullet=4.8\times 10^7\,{\rm M_{\odot}}$,
$\Gamma_K=4.1$, $b_{in}= 1.10 $ and $b_{out}= 1.05$ obtained from the fit of the
second moment and $\kappa=0.9$. Top right: residuals in ${\rm km\,s^{-1}}$.
Bottom right: errors in ${\rm per\,cent}$. Bottom left: linear cuts across the
galaxy centre along the galaxy major axis. Bottom central: linear cuts
across the galaxy centre in a diagonal direction.}
\label{fig:model_V}
\end{figure*}

Adopting the above assumption, we modelled the velocity first moment
using the best-fitting parameters of model E. A comparison between the observed 
rotation velocity field and the modelled first moment for this model is
presented 
in Fig.\,\ref{fig:model_V}.

The top left panel shows the measured velocity field for $V$. The top central
panel shows the modelled velocity first moment $\langle \upsilon_{los}\rangle$
with $\kappa_k = 0.92$ and the best-fitting values of model D. The top right
panel shows the residuals. The bottom right panel shows the residuals in
${\rm per\,cent}$. The bottom left and central panels compare the measured 
and modelled velocity fields along the galaxy major axis  and along a diagonal cut.
It can be observed that the maximum residuals reach about $20\,{\rm km\,s^{-1}}$ which
correspond  to $\approx 50\,{\rm per\,cent}$, as they are observed  in the
regions along the minor  axis where the velocities are lowest. The residuals
are higher than $30\,{\rm per\,cent}$ within the inner $0.5\,{\rm arcsec}$. In
the outer regions the differences between the model and measurements are lower
than  $30\,{\rm per\,cent}$. But the bottom panels highlight that the modelled
velocity first moment does not reproduce the observed rotation mostly in the
region between  $0.1$  and $0.8\,{\rm arcsec}$ (3.5\,pc  and 30\,pc). 

The discrepancy between the modelled velocity first moment field and the
measured centroid velocity field, does not affect our conclusion about the mass
 of the SMBH, that is mostly based on the fit of the second moment. The first
moment derived from the model depends on an extra assumption:  that the
tangential anisotropy is the same everywhere, what may not be true. A more
likely explanation for the discrepancy is however the inaccuracy in the
deprojection of the luminous stellar density from the observed surface
brightness. The deprojection is known to be mathematically degenerate
\citep{Rybicki1987, Gerhard1996}, with the effect becoming quite significant at
low inclination \citep{Romanowsky1997}. Our galaxy is not seen close to edge-on
and could easily hide a weak nuclear disc which would be invisible in the
projected photometry, could be showing up in the velocity field, leading to the
observed discrepancy between the model and the data. This statement is
based on numerical experiments where we tried to force the model to have a flat
disk by making the Gaussians of the MGE to have the same axial ratio. Only in
this way one can lower the model inclination and make the inner disk
intrinsically very thin within the NIFS FoV. With this forced MGE one can better
reproduce the fast increase in rotation and the difference in V$_{rms}$ between
the major and minor axis. Unfortunately the galaxy overall is not well
reproduced by an MGE with constant projected ellipticity. For this reason our
test only shows that the existence of a very thin disk within the NIFS FoV would
improve the fit to the kinematics.

\section{Summary and conclusions}\label{sec:conclusion}

We have presented a two-dimensional mapping of the stellar kinematics of the
inner $3.0\times3.0\,{\rm arcsec^2}$ ($\sim100\,\times100\,{\rm pc^2}$) of the
galaxy NGC\,4258 using NIFS \textit{K}-band data with a signal-to-noise ratio
higher than 50 over most of the observed field, spectral resolution of 5300
and spatial resolution of $\approx$\,4\,${\rm pc}$, allowing to resolve the 
radius of influence of the SMBH ($\approx$\,15\,pc).  The centroid velocity 
field presents a rotational pattern with a maximum velocity of
$\pm80\,{\rm km\,s^{-1}}$, with the SE side approaching and the NW side
receding. The stellar velocity dispersion presents an abrupt increase within 
the inner $0.3\,{\rm arcsec}$ (10\,pc), reaching a value of $180\,{\rm km\,s^{-1}}$
at the nucleus, consistent with the presence of a SMBH there.

In order to model the stellar kinematics we built a Multi-Gaussian Expansion
Jeans Anisotropic Dynamical Model. In this model the velocity second moment
$\langle \upsilon^2_{los} \rangle^{1/2}$ is a function of three free parameters:
the galaxy mass-to-light ratio $\Gamma_K$, the anisotropy in the velocity
$b=\dfrac{\overline{\upsilon_R^2}}{\overline{\upsilon_z^2}}$ and the
mass of the SMBH $M_\bullet$. We performed a $\chi^2$ minimization in this
space of parameters in order to search for the values that
best reproduce the measured $V_{rms}=\sqrt(V^2+\sigma^2)$. In order to
ensure that all regions of the galaxy, independently of its radial position,
have approximately the same relevance in the minimization process we have
weighted the $\chi^2$ minimization by assigning errors to the kinematic
measurements that are inversely proportional to the galaxy surface brightness at
each location. Without this weighting the best-fitting model does not reproduce
the measured $V_{\rm rms}$ in the nuclear region of the galaxy and provides a
a wrong (too high) value for the mass of the SMBH.

We have tried models with only one value for the velocity anisotropy and
models with two values. The best-fitting model was obtained adopting a galaxy
inclination $i=64$\textdegree\, and considering that the galaxy has a radially 
variable anisotropy in the velocity second moment, being the values of the 
anisotropy parameter for the inner 4 Gaussians $b_{in}=1.10\, (\beta_z = 0.09)$ 
and for the remaining 8 outer Gaussians $b_{out}=1.05\, (\beta_z = 0.05)$.

Considering the 3$\sigma$ confidence intervals, we have obtained a
mass-to-light ratio of $\Gamma_k = 4.1^{+0.4}_{-0.5}$ and the best-fitting
SMBH mass of $M_\bullet = 4.8^{+0.8}_{-0.9}\times 10^7\,{\rm M_\odot}$.  These
3$\sigma$ uncertainties of our model are comparable to the typical values
for the uncertainties of stellar dynamical models for other galaxies, that
usually are lower than $50\,{\rm per\,cent}$ for the mass of the SMBH
\cite{McConnell2011}. This provides further confirmation of the robustness of
the stellar dynamical determinations. It is worth noting that an accurate BH
masses was obtained even when using a simple model and only fitting the
kinematics within the innermost few arcseconds, without the need for a
large-scale model of the galaxy dynamics.

\section*{ACKNOWLEDGEMENTS}
Based on observations obtained at the Gemini Observatory, which is
operated by the Association of Universities for Research in Astronomy,
Inc., under a cooperative agreement with the NSF on behalf of
the Gemini partnership: the National Science Foundation (United
States), the Science and Technology Facilities Council (United
Kingdom), the National Research Council (Canada), CONICYT
(Chile), the Australian Research Council (Australia), Minist\'erio da
Ci\^encia e Tecnologia (Brazil) and SECYT (Argentina). This work
has been partially supported by the Brazilian institution CNPq. MC acknowledges
support from a Royal Society University Research Fellowship. We thank the
referee, Remco van den Bosch for the valuable suggestions that helped to improve
the paper.

\newpage

\begin{appendix}

 \section{MGE Parameters for the NICMOS and NIFS PSF}\label{app:psf_models}
 In this appendix we present the MGE models for the NICMOS and NIFS PSF's and
the agreement of the MGE model of the NICMOS PSF with the on of the {\sc tiny
tim}  model. 

\subsection{The NICMOS PSF}\label{app:psf_nic}
\begin{figure}
  \centering 
\includegraphics[width=0.5\textwidth]{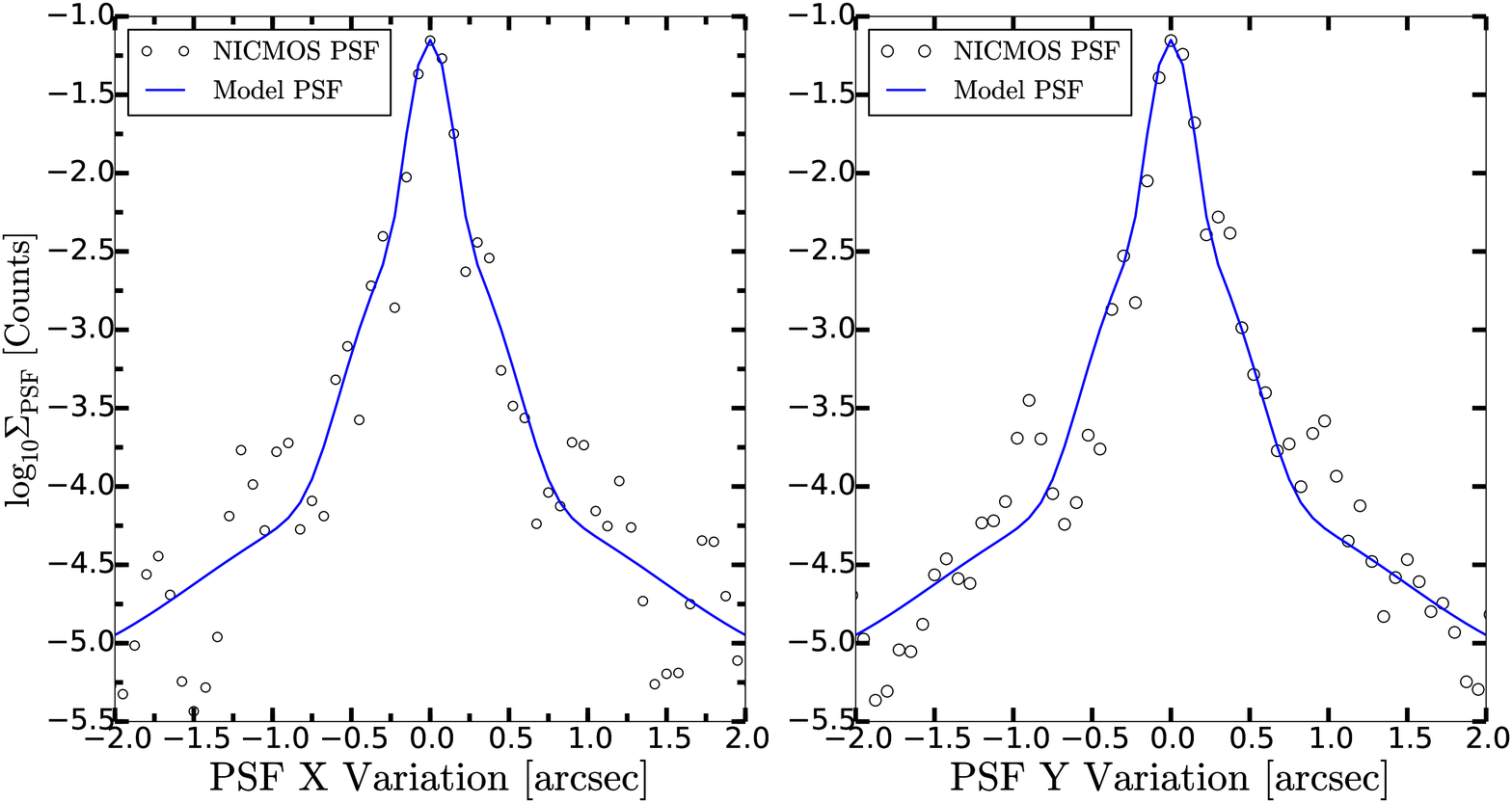}
\caption{NICMOS PSF fit. Left: linear cut along the detector horizontal
direction showing the radial profile of the NICMOS PSF of the {\sc tiny tim}
model (black open circles) and of the resulting MGE model for the PSF (blue
solid line). Right: Same as the left but for the detector vertical direction.}
\label{fig:psf_nic}
\end{figure}

We modelled the NICMOS PSF by a set of four approximately circular Gaussians
using the MGE method. The resulting Gaussian parameters are presented in
Tab.\,\ref{tab:gauss_ncmos}. 
\begin{table}
\centering
\caption{Gaussian parameters of the MGE model for the NICMOS PSF.}
\label{tab:gauss_ncmos}
\begin{tabular}{@{}ccc@{}}
\hline
$\Sigma'_k {\rm [Total\,Counts]}$ &$ \sigma'_k\,[arcsec] $ & $q'_k$ \\
\hline    
 $ 0.52  $& $ 0.084 $ &$ \approx 1.0 $  \\ 
 $ 0.34  $& $ 0.242 $ &$ \approx 1.0 $  \\ 
 $ 0.06  $& $ 0.789 $ &$ \approx 1.0 $  \\ 
 $ 0.08  $& $ 2.671 $ &$ \approx 1.0 $  \\
\hline                     
\end{tabular}  
\end{table} 
In Fig.\,\ref{fig:psf_nic} we present the comparison between the luminosity
profiles of the {\sc tiny tim} PSF showed by the black open circles and our MGE
modelled PSF showed by the blue solid line. 

\begin{figure} 
  \centering 
\includegraphics[width=0.5\textwidth]{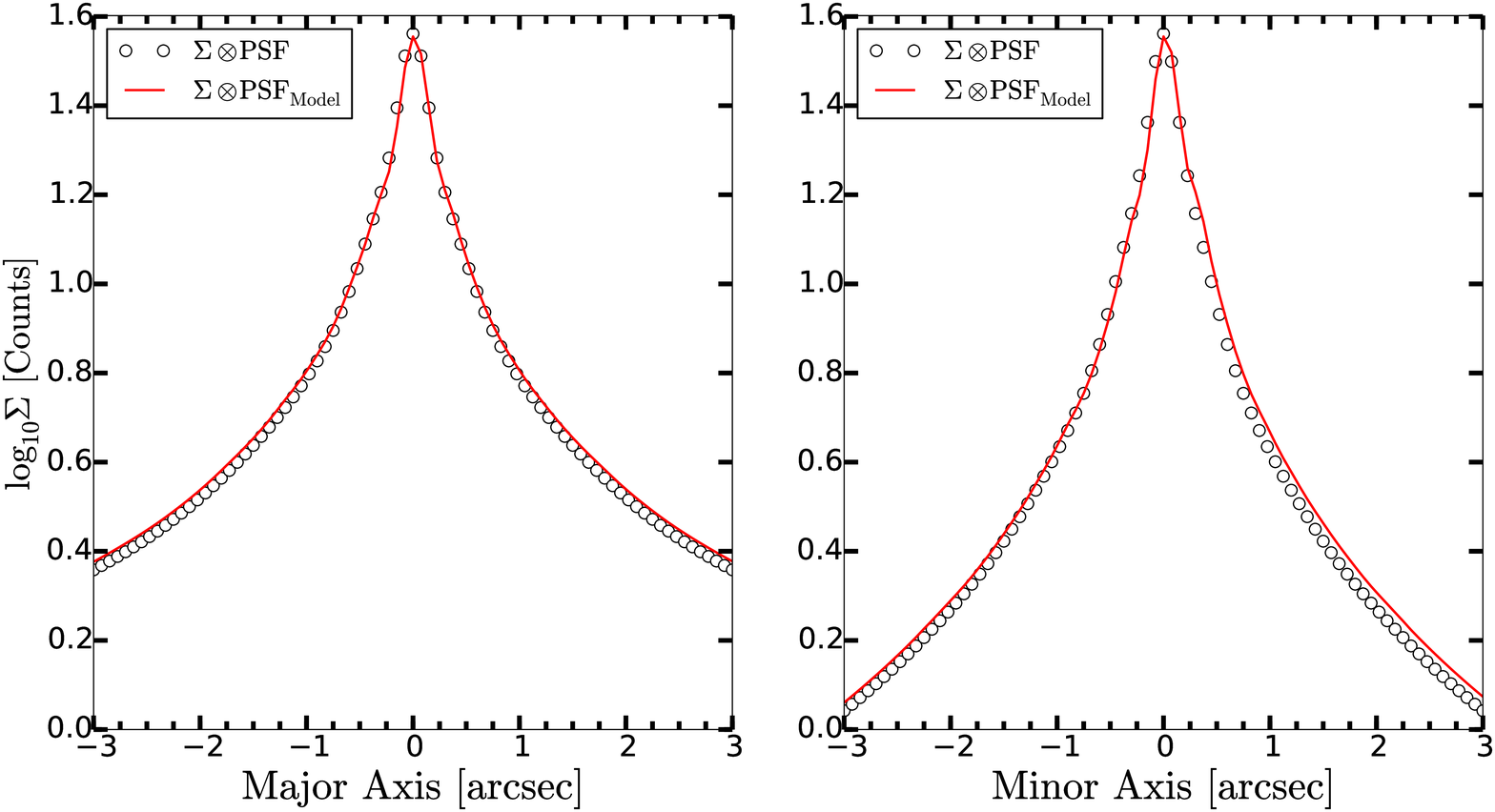}
\caption{Comparison of the convolution of the MGE model for the surface
brightness distribution with the {\sc tiny tim} PSF and with the MGE model for
the PSF. Left: cuts along galaxy major axis. Right: cuts along galaxy minor
axis.}
\label{fig:psf_convolution}
\end{figure}

In order to verify the effects of the small differences between the MGE model
and the real PSF in the convolution procedure we present in
Fig.\,\ref{fig:psf_convolution} a comparison of the convolution of
the surface brightness distribution of NGC\,4252 presented in
sec.\,\ref{sec:mge} with the {\sc tiny tim} PSF (black open circles) and with
our MGE
model for the NICMOS PSF (red solid line) for the central $6.0\,{\rm arcsec}$ of
the galaxy. The left panel shows a linear cut across the galaxy centre and along
the photometric major axis and the right panel along the galaxy minor axis. In
both directions the differences are very small in most of the regions.

\subsection{The NIFS PSF}\label{app:psf_nifs}

\begin{figure} 
  \centering 
\includegraphics[width=0.5\textwidth]{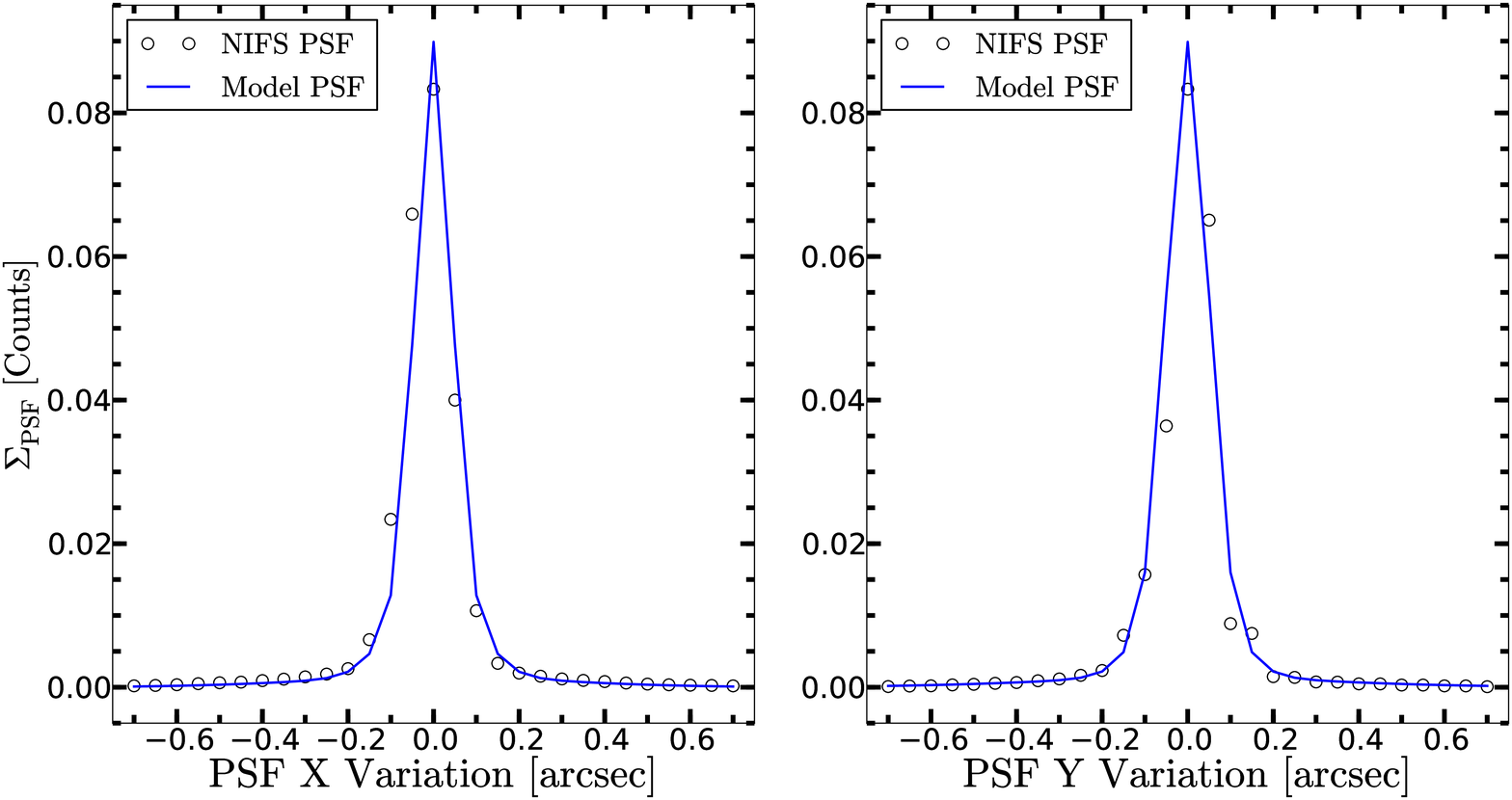}
\caption{NIFS PSF fit. Left: linear cut along the detector horizontal
direction showing the radial profile of the NIFS PSF obtained from observations
of three stars (black open circles) and of the resulting MGE model for the PSF
(blue solid line). Right: Same as the left but for the detector vertical
direction.}
\label{fig:psf_nifs}
\end{figure}

The MGE model of the NIFS PSF is performed using the average image of three 
reconstructed images from the data cubes of three stars observed with NIFS in
the same night of the observations of the galaxy. The Gaussian parameters of
the MGE model for the NIFS PSF are presented in Tab.\,\ref{tab:gauss_nifs}.
\begin{table}
\centering
\caption{Gaussian parameters of the MGE model for the NIFS PSF.}
\label{tab:gauss_nifs}
\begin{tabular}{@{}ccc@{}}
\hline
$\Sigma'_k {\rm [Total\,Counts]} $ &$ \sigma'_k\,[arcsec] $ & $q'_k$  \\
\hline    
 $ 0.29  $& $ 0.045 $&$ \approx 1.0 $  \\
 $ 0.20  $& $ 0.069 $&$ \approx 1.0 $  \\
 $ 0.15  $& $ 0.103 $&$ \approx 1.0 $  \\
 $ 0.27  $& $ 0.317 $&$ \approx 1.0 $  \\
 $ 0.09  $& $ 0.543 $&$ \approx 1.0 $ \\
\hline                     
\end{tabular}  
\end{table} 
In Fig.\,\ref{fig:psf_nifs} we present a comparison between the resulting MGE
model and the NIFS PSF. In the left panel we show a linear cut along the 
detector horizontal axis of the NIFS PSF (black open circles) and of the MGE
model (blue solid line). In the right panel we show a linear
cut along the detector vertical axis.

\section{Comparison between the kinematics measurements of NIFS  and STIS
}\label{app:kinematics_stis} 
\begin{figure} 
  \centering 
\includegraphics[width=0.48\textwidth]{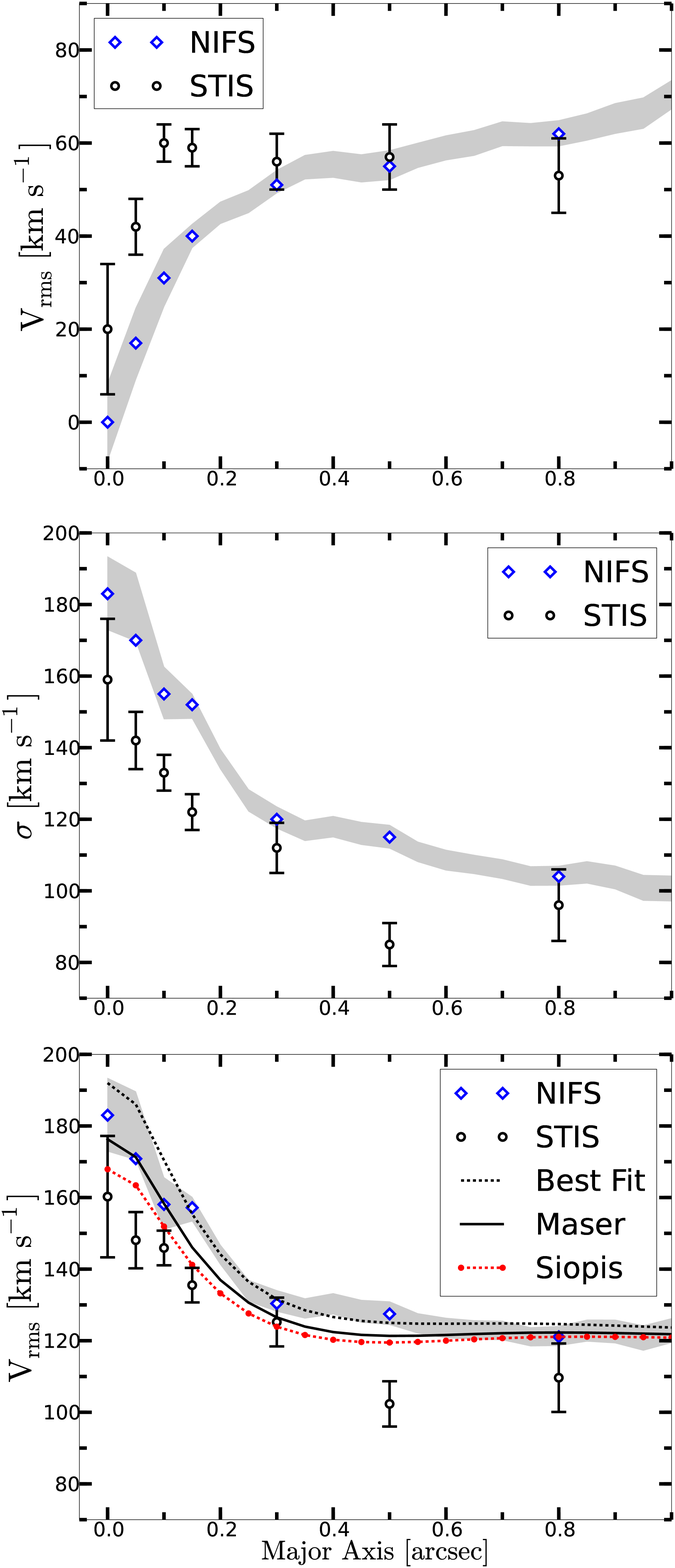}
\caption{Comparison between STIS and NIFS kinematics along the galaxy major
axis. Top: The blue diamonds are the measures of the centroid velocities from
our NIFS data with the 1$\sigma$ error bar represented by the shaded region,
the black circles with error bars are the measurements from STIS observations.
Middle: Velocity dispersions from the NIFS and STIS observations. Bottom: The
blue diamonds and black circles are the NIFS and STIS measured V$_{\rm rms}$
respectively. The dashed line is our best-fitting model for the velocity second
moment, the continuous line is the modelled velocity second moment using the
mass of the SMBH from the maser determination and the dashed-dotted line is the
model using the mass of the SMBH from \citet{siopis09}. }
\label{fig:kin1}
\end{figure}
In this section we present a comparison between the measured velocities
and velocity dispersions of our NIFS data and the previous published long-slit
data from the HST/STIS instrument used by \citet{siopis09} to determine the mass
of the SMBH in the nucleus of NGC\,4258 through Schwarzschild dynamical orbit
superposition. In Tab.\,\ref{tab:kin} we present kinematic measurements with
1$\sigma$ errors from the STIS data \citep{siopis09} for the positions along
the galaxy major axis coincident with our NIFS measurements. In column one we
list the positions of the extractions; in column two, the values of the centroid
velocities obtained from the STIS data; in column three, the values of the
centroid velocities from the NIFS data; in columns four and five we present
the values of the velocity dispersion measurements from the STIS  and NIFS data
respectively.

\begin{table*}
\begin{center}
\caption{Comparison between NIFS and STIS kinematic measurements along the
galaxy major axis.}
\label{tab:kin}
\begin{tabular}{@{}ccccc@{}}
\hline
$x'\,[arcsec]$ & $V_{\rm STIS}\,[km \,s^{-1}] $ & $V_{\rm NIFS}\,[km \,s^{-1}]
$&    $ \sigma_{\rm STIS}\,[km \,s^{-1}] $ & $ \sigma_{\rm NIFS}\,[km \,s^{-1}]
$ \\
\hline    
$ 0.00$ & $ 20\pm 14$ &  $ 0\pm8   $ & $ 159\pm17   $ & $ 183\pm10  $\\
$ 0.05$ & $ 42\pm 6 $ &  $ 17\pm8   $ & $ 142\pm8  $ & $ 170\pm10  $\\
$ 0.10$ & $ 60\pm 4 $ &  $ 31\pm6   $ & $ 133\pm5  $ & $ 155\pm7  $\\
$ 0.17$ & $ 59\pm 4 $ &  $ 40\pm3   $ & $ 122\pm5  $ & $ 152\pm3  $\\
$ 0.30$ & $ 56\pm 6 $ &  $ 51\pm2   $ & $ 112\pm7  $ & $ 120\pm3  $\\
$ 0.50$ & $ 57\pm 7 $ &  $ 55\pm3   $ & $ 85\pm 6  $ & $ 115\pm3  $\\
$ 0.80$ & $ 53\pm 8 $ &  $ 62\pm3   $ & $ 96\pm 10  $ & $ 104\pm3  $\\

\hline                     
\end{tabular}
\end{center}
\begin{small}
Note: Column 1 are the positions of the extractions along the galaxy major axis;
Column 2 are the centroid velocities from STIS observations as tabulate in Table
3 of \citet{siopis09}; Column 3 are the centroid velocities from the NIFS
observations; Column 4 are the velocity dispersions from the STIS observations;
Column 5 are the velocity dispersions from the NIFS observations.
\end{small}
\end{table*} 

The differences between the measured velocities from both instruments
are shown in Fig.\,\ref{fig:kin1}. The top panel shows the centroid velocities:
the blue diamonds are the NIFS data with the 1$\sigma$ errors being represent by
the shaded grey band, the open circles with the error bars are the STIS data. In
the region inside 0.2 arc seconds the STIS data shows a steeper velocity curve
than the NIFS data, while for the outer region the error bars overlap. We
attribute this difference to the somewhat better angular resolution of the STIS
when compared to that of NIFS.

The middle panel shows the velocity dispersions: the NIFS data have
$\sigma$ values $\approx${\rm \,20\, km\, s{$^{-1}$}} higher than those from the
STIS data. We attribute this difference to the smaller {\rm FWHM} of
the  STIS PSF and to the different methods and templates used for the $\sigma$ 
determinations: while we have used the stellar templates library  of
\cite{winge09}, \citet{siopis09} used only one stellar template (HR\,7615).

In the bottom panel we show the resulting V$\rm_{rms}$ compared to
different models. The black dashed line is the velocity second moment along the
galaxy major axis from our best-fitting model with the parameters
$b_{in}=1.10\, (\beta_z = 0.09)$, $b_{out}=1.05\, (\beta_z = 0.05)$,
$\Gamma_K=4.1$ and $M_{\bullet}\, = 4.8\times 10^7\,{\rm M_\odot}$; the black
continuous line is the second moment from a model with the same anisotropy and
mass-to-light ratio of the best-fitting model but using the value of
$M_{\bullet\,{\rm Maser}} = 3.82\times 10^7\,{\rm M_\odot}$; and the red
dashed-dotted line is a model using the value of $M_{\bullet\,{\rm Schw}} =
3.3\times 10^7\,{\rm M_\odot}$. While our best-fitting model gives a $M_\bullet$
somewhat larger than that of the maser determination, the \citet{siopis09} 
gives a  $M_\bullet$ somewhat smaller, but the NIFS measurements gives
V$\rm_{rms}$  values closer to the one from the model with
$M_\bullet$\,=\,$M_{\bullet\,{\rm Maser}}$.

\section{Subtraction of the point source}\label{app:oversubtraction}
\begin{figure*} 
  \centering 
\includegraphics[width=0.9\textwidth]{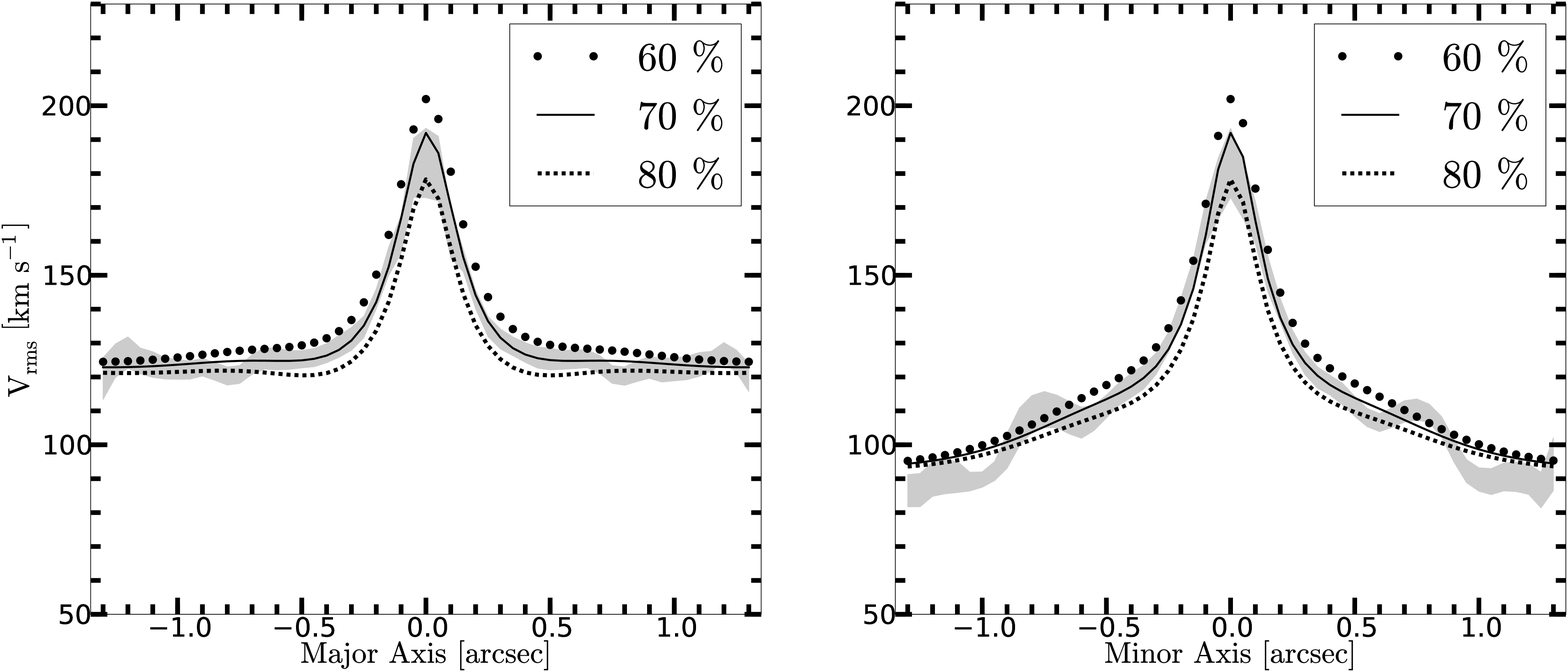}
\caption{Effects of the point source subtraction on the modelled second
moment - Test 1. Left panel: The continuous line is a cut along the galaxy major
axis of our best-fitting model, the dotted line shows the modelled second moment
using the the same dynamical parameters of our best-fitting model and
subtracting 10 per cent ($\approx$\,0.27\,$\times$\,10$^7$\,L$_\odot$) more
luminosity in the AGN contribution. The black fil circles represent the
resulting model subtracting 10 per cent less luminosity in the AGN subtraction.
In the two cases the luminosity is subtract from the innermost gaussian of the
MGE model. The shaded band are the 1$\sigma$ confidence intervals for the
V$_{rms}$. Right panel: Same as the left panel but for the galaxy minor axis.}
\label{fig:psf_subtraction1}
\end{figure*}

\begin{figure*} 
  \centering 
\includegraphics[width=0.9\textwidth]{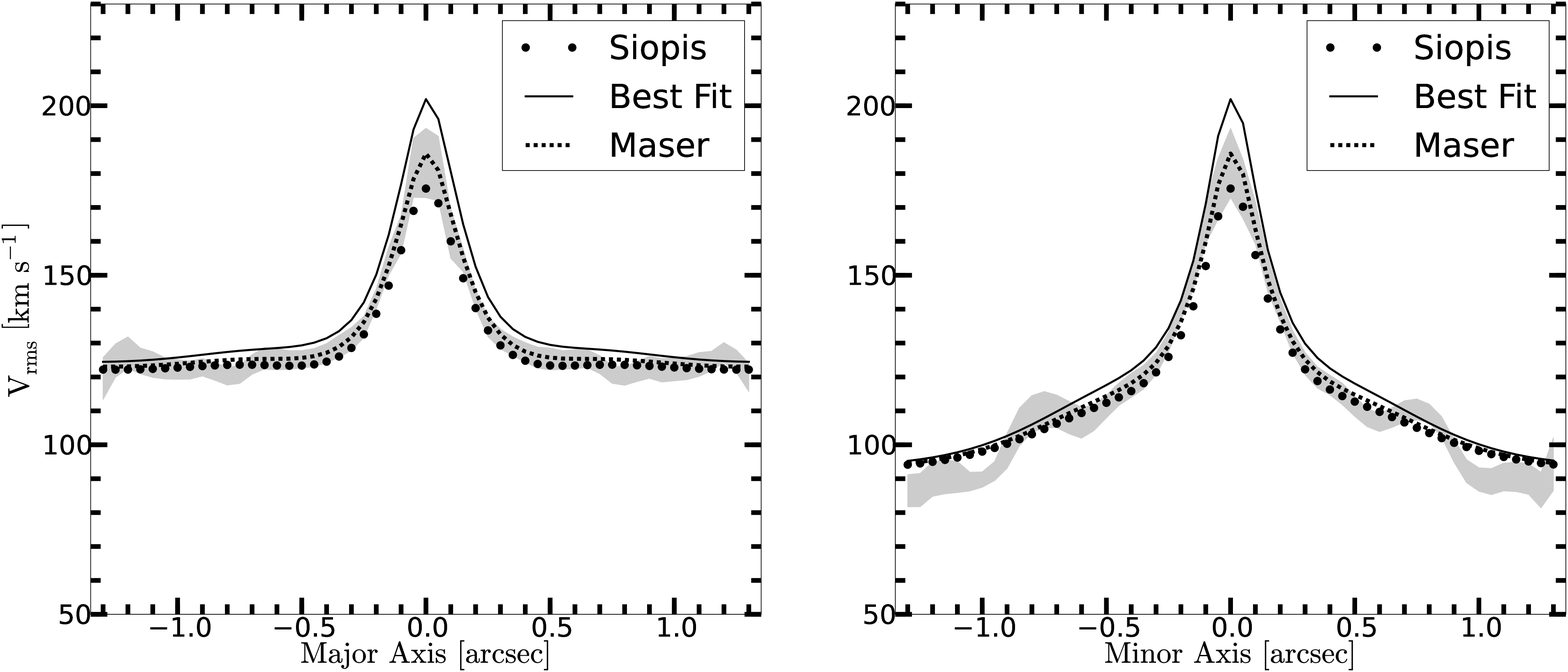}
\caption{Effects of the point source subtraction on the modelled second moment
- Test 2. Left: the continuous line is a cut along the galaxy major axis showing
the values of the velocity second moment for the parameters of our
best-fitting model and subtracting 10 per cent less luminosity
($\approx$\,0.27\,$\times$\,10$^7$\,L$_\odot$ ) in the AGN contribution, The
dotted line corresponds to the model using the value of the mass of the SMBH of
the maser determination (3.82\,$\times$\,10$^7$\,M$_\odot$) and the black
bullets is the model using the value of the mass of the SMBH from the previous
stellar dynamical determination of \citet{siopis09}
(3.3\,$\times$\,10$^7$\,M$_\odot$).The shaded band are the 1$\sigma$ confidence
intervals for the V$_{rms}$.}
\label{fig:psf_subtraction2}
\end{figure*}

In section\,\ref{sec:photodata}, we have determined that the luminosity
of the AGN contributes with approximately 70 per cent to the total intensity of
the central pixel of the NICMOS image. Thus the total luminosity subtracted from
the galaxy image in order to eliminate the contribution of the AGN is of the
order of $\approx$\,1.9\,$\times$10$^7\,{\rm L_{\odot K}}$. Using the
mass-to-light ratio of $\Gamma_k$ = 4.1 as obtained from our best-fitting model
this corresponds to a mass of $\approx$\,7.8\,$\times$10$^7\,{\rm M_\odot}$. In
order of investigate how a possible over- or under-subtraction of the AGN
contribution affects the determination of the mass of the SMBH we performed two
simplified tests. 

In the first test we assume that only the innermost gaussian component
of the MGE model is affected by the subtraction of the point source. This is a
plausible assumption as all the other Gaussians components of the model extend
beyond the first Airy ring of the PSF and after the convolution
of the surface brightness distribution with the PSF the Gaussian components
become even flatter. We then added and subtracted 
$\approx$\,0.27\,$\times$10$^7\,{\rm L_{\odot K}}$
(or $\approx$\,1.12\,$\times$10$^7\,{\rm M_\odot}$ ) from the innermost
gaussian. In practice this is equivalent to instead of subtracting 70 per cent
of the intensity of the central pixel to subtract  60 per cent and 80 per cent
respectively. The effects on the modelled velocity second moment for the
three different subtractions of the AGN contribution are shown in
Fig.\,\ref{fig:psf_subtraction1}, that presents cuts along the galaxy
major and minor axis showing the variation of the modelled velocity second
moments. The continuous line is our original best-fitting model (Model D),
the dashed line is a model with the same dynamical parameters of the
best-fitting model but with a additional mass of
$\approx$\,1.12\,$\times$10$^7\,{\rm M_\odot}$ to the central gaussian, and the
shaded band is the 1$\sigma$ confidence region for the kinematic measurements.
The black filled circles are the model where we subtracted the same mass
from the innermost gaussian. The resulting variation in the velocity second
moment of the central pixel of the galaxy for this variation in the
mass is of the order of $\pm$12\,${\rm km\,s^{-1}}$.

In the second test we assumed that the innermost gaussian has a
luminosity of  $\approx$\,0.27\,$\times$10$^7\,{\rm L_{\odot K}}$ higher, (this
corresponds to subtract only 60 percent from the intensity of central pixel,
this test is motivated by the fact that this amounth of light corresponds
approximately to a stellar mass that is equal to the difference in the values
for the mass of the SMBH obtained from our best-fitting model and the value from
the maser determination) and we use the parameters of anisotropy and
mass-to-light ratio of our best-fitting models to investigate how well the three
different determinations for the mass of the SMBH (Maser, STIS, ours) reproduce
the observed kinematics. The resulting modelled velocity second moments along
the galaxy major and minor axiz are presented respectively in the left and right
panels of Fig.\,\ref{fig:psf_subtraction2}. The continuous line corresponds to
our best-fitting model, the dashed line is the model using the value for the
mass of the SMBH from the maser determination and the black filled circles are
the model with the value for the mass of the SMBH determined by
\citet{siopis09}. The shaded band is the 1$\sigma$ confidence region for
the kinematic measurements. Under these assumptions, the best-fit is obtained
for the model with $M_{\bullet\,{\rm Maser}} = 3.82\times 10^7\,{\rm M_\odot}$.
The model using a mass of $M_{\bullet\,{\rm STIS}} = 3.3\times 10^7\,{\rm
M_\odot}$ for the SMBH shows a poor but a similar fit to the data to that of
our model only  that our model are close to the upper envelope of the 1$\sigma$
band while the of \citet{siopis09} are close to the lower envelope.

But we would to point out that the uncertainty in the determination of
the AGN is snaller than 10 per cent. Actualy even an over-subtraction of
$\approx$\,5\, per cent already leaves a signature on the NICMOS PSF in the
image. Our choice of $\approx$\,5\, per cent is because it corresponds im mass
to the difference between the value of the SMBH mass of our best-fitting model
and that of the maser determination.

\end{appendix}


\end{document}